\documentclass[10pt]{article}

\usepackage{graphicx,subfig}
\usepackage{times}
\usepackage{tabu,booktabs,multirow}
\usepackage{mathtools}
\usepackage{enumitem}
\usepackage{url}

\usepackage{pbltx}

\newcommand{\xitem}{\textbullet}
\newcommand{\shortcite}[1]{\cite{#1}}

\begin{document}

\title{Analyzing Regrettable Communications on Twitter:
Characterizing Deleted Tweets and Their Authors}

\author{Parantapa Bhattacharya, Saptarshi Ghosh and Niloy Ganguly\\
\{parantapa,saptarshi,niloy\}@cse.iitkgp.ernet.in}

\maketitle


\begin{abstract}

Over 500 million tweets are posted in Twitter each day,
out of which about 11\% tweets are deleted by the users posting them.
This phenomenon of widespread deletion of tweets leads to a number of questions:
what kind of content posted by users
makes them want to delete them later?
Are users of certain predispositions more likely
to post regrettable tweets, deleting them later?
In this paper we provide a detailed characterization of
tweets posted and then later deleted by their authors.
We collected tweets from over 200 thousand Twitter users
during a period of four weeks.
Our characterization shows significant personality differences
between users who delete their tweets and those who do not.
We find that users who delete their tweets are more likely to be
extroverted and neurotic while being less conscientious.
Also, we find that deleted tweets
while containing less information
and being less conversational,
contain significant indications of regrettable content.
Since users of online communication do not have instant social cues
(like listener's body language)
to gauge the impact of their words,
they are often delayed in employing repair strategies.
Finally, we build a classifier which takes
textual, contextual, as well as user features
to predict if a tweet will be deleted or not.
The classifier achieves a F1-score of 0.78 and the precision increases when we consider response features of the tweets.%
\footnote{%
This work is an extended version of the short paper:
P. Bhattacharya and N. Ganguly,
``Characterizing Deleted Tweets and Their Authors'',
In Proceedings of the 10th AAAI International Conference on
Web and Social Media (ICWSM), pp 547--550, 2016.

This study was conducted during the year 2016
when Parantapa Bhattacharya was a student at 
Indian Institute of Technology Kharagpur.
The datasets collected for these studies have since been permanently deleted.

This study was conducted respecting the guidelines
set by our institute's ethics board
and with their knowledge and permission.}

\end{abstract}

\section{Introduction}
\label{sec:intro}

With the increasing popularity of 
Online Social Networks (OSNs), millions of users
take resort to these media for sharing their thoughts and ideas
with hundreds and thousands of others.
However, in doing so, the OSN users have
also become susceptible
to inadvertently exposing
potentially private and embarrassing information.
Unwittingly exposing personal information and opinions
can have wide ranging negative consequences --
from harmless awkwardness
to much more severe ones
such as losing jobs%
\footnote{\url{http://www.businessinsider.com/twitter-fired-2011-5?IR=T}}
and even getting prosecuted%
\footnote{\url{http://www.bbc.com/news/world-middle-east-30903294}}%
\footnote{\url{http://www.theguardian.com/uk-news/2014/oct/20/man-jailed-antisemitic-tweet-labour-mp}}.

Once a user realizes that she regrets a post that she has made,
the repairing strategy that is most commonly adopted
is to delete the offending post~\cite{wang-soups11,sleeper-chi13}.
We observe that over 11\% of tweets created,
are deleted either by Twitter or the users posting them.

In light of the substantial portions of content
being regularly removed from Twitter by their authors,
a number of interesting questions arise naturally --
(1)~Are all Twitter users equally likely to delete their tweets,
or is it that only users of particular predispositions
account for the majority of tweet deletions?
(2)~Can distinctive characterizations be obtained
for tweets that are deleted in the future,
compared to tweets that are not deleted by their authors?
(3)~Do responses from other users
lead the author of a tweet
into realizing the potential impact of their tweet,
prompting them to delete it?

For researchers and developers, it is fundamentally important
to understand the above issues,
if they are to build systems that help users
in managing potentially regrettable posts.
A user's personality, her desired public image,
and her general writing style
overwhelmingly determine what she finds regrettable.
Whether she chooses to delete one of her own posts or not
is heavily dependent on her personal preferences.
Understanding these factors and taking them into account
is essential for every privacy conscious system.

There have been some previous works \cite{wang-soups11,sleeper-chi13}
which mainly relied on user surveys
to understand these issues.
While such first hand accounts
provide useful insights into the process of tweet deletion,
it is difficult to directly utilize them to build systems,
that help users manage potentially regrettable posts.
Further, when a user makes a regrettable remark
in an offline conversation,
her audience's response, especially body-language,
lets her quickly gauge the situation.
This missing aspect of interaction
often delays users
from realizing their errors
in online communications~\cite{sleeper-chi13}.
Existing studies of deleted tweets~\cite{almuhimedi-cscw13,zhou-www16}
have generally ignored this aspect of communication
when trying to understand reasons behind tweet deletions.


With an eventual aim to build a classifier
for early identification of potentially regrettable tweets,
this paper tries to answer the following research questions:
\begin{itemize}
  \item Q1: Do deleted tweets demonstrate significantly higher
    indications of regrettable content?
    (Section~\ref{sec:delcategorize})
  \item Q2: Does there exist significant differences in personality,
    between users who delete their tweets and those who do not?
    (Section~\ref{sec:delusers})
  \item Q3: Are there significant linguistic differences
    between the deleted and non-deleted tweets of the same user?
    (Section~\ref{sec:delcontent})
  \item Q4: Does the sentiment of replies
    to tweets that are deleted in the future,
    show significant variations
    when compared to replies to tweets
    that are not deleted?
    (Section~\ref{sec:response})
  \item Q5: With what degree of confidence can a classifier predict
    that a tweet will be deleted or not?
    (Section~\ref{sec:classify})
\end{itemize}
In this paper,
we present a large scale empirical study of all tweets
posted and deleted by over 200 thousand users
during a four week period in August 2015,
to answer the questions presented above.

Note that, in their study of tweet deletions,
Almuhimedi et al.~\cite{almuhimedi-cscw13} reported
to being unable to find differences in
``distributions of stereotypically regrettable topics''
due to the ``sheer volume and variety of reasons'' for which
tweets are deleted.
To alleviate this problem, we undertook 
a systematic data cleaning procedure,
to remove malicious users, automated tweets, superficial deletion,
and deleted retweets from our dataset
(Section~\ref{sec:dataset}).

In Sections \ref{sec:delusers} and \ref{sec:delcontent} we present
personality based characterizations of
users who are more likely to delete their tweets,
and demonstrate detailed linguistic differences
between deleted and non-deleted tweets of same users.
The differences discovered in both cases,
were significantly facilitated
by the rigorous data cleanup procedure
undertaken for this work.

Earlier user survey based studies~\cite{wang-soups11,sleeper-chi13},
that analyzed regrettable content on Twitter,
had found that regrets in online communication can be categorized
into a small set of classes such as,
``Direct attack'', ``Direct criticism'', ``Implied Criticism'', etc.
To understand what percentage of deleted tweets actually
fit into those classes, we performed a manual annotation study,
whereby 100 deleted and 100 non-deleted tweets were manually
annotated by human annotators into those classes.
Surprisingly, we found that the annotators agreed
that only 16 out of the 100 deleted tweets looked regrettable,
while 6 out of 100 non-deleted tweets also looked regrettable.
While this difference in statistically significant
(using Fisher's Exact test with Odds Ratio=$0.33$ and $p=0.04$)
it seems that for the majority of deleted tweets (84\%)
the personal nature of the underlying regrettable cause,
which triggered the deletion of the tweet,
is difficult to uncover for third party viewers,
from just the content of the tweets.

While it may be difficult for third party human annotators
to infer the underlying causes of regret
which triggered the deletion of a specific tweet,
there still may exist systematic characteristic
in tweets that are deleted,
that can be helpful in identifying them early.
Thus, we undertook an analysis,
trying to identify the salient personality traits
of users who are more likely to delete their tweets,
as well as uncovering detailed linguistic differences
between tweets that are deleted and those that are not.



We found that users who are more likely to delete their tweets
show systematic differences in their Big-Five
personality characteristics~\cite{big-five}
compared to non-deleters;
they are more likely to be extroverted and neurotic
while being less conscientious.
\Replacement{Further, we also found that deleted tweets
generally contain less informative content
and are also less conversational.
Surprisingly, we found that deleted tweets show definite signs
of being thoughtfully constructed
although use of profanity and swear words
is significantly higher in them,
compared to tweets that are not deleted.}
Additionally, deleted tweets are less likely to get responses,
but when they do, they get significantly more negative replies.

\todoNG{too many features!!}
Utilizing the insights on personality characteristics of users
and linguistic characteristics of tweets,
we were able to build a classifier
which uses textual, contextual, as well as user features
to distinguish between deleted and non-deleted tweets of users,
with a F1-score of 0.78.
The classifier was built in two stages
to effectively utilize high-dimensional but sparse text features,
as well as low dimensional and dense contextual features and user features.
Interestingly, we find that user features,
despite not having any direct information about the tweet,
are the most discriminating features
for predicting tweet deletion probabilities.
We postulate that they do so by indirectly encoding
signals about the author's personality.

\section{Related Work}
\label{sec:related}
\todoNG{I am not sure whether opinion can be plural}

The Twitter microblogging network
is one of the most popular
online social networking platforms
with over 500 million tweets posted per day~\cite{500M-tweets-daily}.
Twitter has been very popular
among social networks researchers.
A recent work by Liu et al.~\cite{liu-icwsm14}
chronicles the evolution of Twitter over the years.

Users in Twitter post tweets,
which are microblogs
limited to 140 characters in length.
In Twitter, users follow each other to receive tweets.
If user A follows user B,
user A receives all tweets posted by user B.
Here A is termed as a \emph{follower} of user B.
While B is termed as a \emph{followee} of user A.

For this work,
we distinguish three types of tweets:
regular \emph{tweets}
that are received by all followers of the user posting them,
\emph{replies} which are tweets characterized by
the screen name of the recipient user at the beginning of the tweet,
and \emph{retweets} which are copies of tweets made by other users.
Retweets, like regular tweets,
are circulated to followers of the user retweeting them,
while replies created by a user are not
received by her followers.

Twitter doesn't provide any option
for editing or changing a tweet.
However, a user can choose to delete her tweets.
Once, a tweet is deleted,
it is no longer visible by anyone.

\subsection{Regrettable communications in the offline world}

Often when speaking to someone,
people say regrettable things
with or without really meaning what they had said.
Knapp et al.~\shortcite{knapp-jc86} studied the problem
of regrets in verbal communications.
They surveyed 155 participants
and came up with a list of eleven major categories of regrets
as reported by the participants.
They found that blunders, direct attacks/criticism,
group reference, and revealing too much
accounted for 73.8\%
of the stated causes of regret
in their surveys.

Meyer in two separate works~\cite{meyer-ccr04,meyer-jlsp11}
studied the context in which
regrettable communications were made.
They also studied
the aftermath of regrettable communications
and repair strategies used by
individuals to remedy the situations.
In the two separates studies,
the authors surveyed 204~\cite{meyer-ccr04}
and 173~\cite{meyer-jlsp11}
undergraduate students at a large midwestern university.
The survey participants were asked about
their emotional state
when making the regrettable communication
and the repair strategies undertaken by them.
These studies concluded that,
the regretted communications were rarely directed
towards more than one person
and negative emotional states were often at fault
when making a regrettable remark.

\subsection{Regrettable communications in online social networks}

Most communications in online social networks
happen in broadcast mode.
This makes communications
reach large audience
quickly and easily.
However, with posts having such large visibility
the problems of regrettable communications multiply.

In Twitter, the primary mechanism for restricting
ones visibility is by making their account protected.
Only the followers of a protected account
can view the tweets posted by the account.
However, Meeder et al.~\cite{meeder-w2sp10}
conducted a study showing the flawed nature of such
a simplistic privacy mechanism.
Tweets from protected accounts can be widely circulated
(beyond the protected user's follower circle)
if it is retweeted by any of her followers.
Such, unintended audience increases the probability
of a user's communications to become regrettable.

The work by Mao et al.~\cite{mao-wpes11}
showed that despite knowing the public nature of social networks,
users often post private and sensitive information online.
Wang et al.~\cite{wang-soups11} presented
a qualitative study on regrettable communications in Facebook.
They observed that
the major causes of regrettable communications in Facebook
were ``centered around
sensitive topics, emotional content, and unintended audience''.

One of the primary methods
by which users
avoid situations of regrettable communications
is via self censorship.
Sleeper et al.~\cite{sleeper-cscw13} performed a survey based study
and concluded that users on Facebook
self-censor primarily to maintain self images
across large audience
that compromises of their many friend circles.
Das and Kramer~\cite{das-icwsm13} performed a large scale empirical study
with 3.9 million users
where they monitored user behavior
when posting content to Facebook.
They concluded that about 71\% of Facebook users self-censor
to avoid regrettable communications.

\subsection{Regrettable communications and deleted tweets in Twitter}


\Replacement{In a survey of 1,221 users,
recruited via Amazon Mechanical Turk,
Sleeper et al.~\cite{sleeper-chi13}
asked participants to describe
``one thing they had said and then later regretted'',
on Twitter and in the offline world.
Using the categorization system
developed by Knapp et al.~\cite{knapp-jc86},
the authors categorized the stated regrets in Twitter
and the offline world into eleven categories.
The authors found that
blunders, direct attacks/criticism,
group reference, and revealing too much
accounted for 83\% of
the stated causes of regret.}

Mondal et al.~\cite{mondal-soups16} in a very recent work,
performed a large scale measurement study
trying to understand how users control the exposure
of their publicly posted content
over longer periods of time.
Interestingly, they found that for all posts that were
published over six years ago,
over 28\% of them are no longer available on Twitter.
Mondal et al. attribute this removal of public content
to privacy conscious users, who have either deleted their
Twitter account, or have made their account protected.
Further, they go on to show that although such content
has been made private by their authors,
significant details about such content can be
inferred using responses to such posts
that are still publicly available.

Preliminary works on deleted tweets
have tried building classifiers
to predict whether a tweet will be deleted or not.
Petrovic et al.~\cite{petrovic-unpub13} collected a 75 million tweet sample
from the Twitter streaming API,
out of which 2.4 million were deleted.
Using an SVM classifier with a linear kernel
and a bag of words feature model,
they achieved a very low F1-score of 0.27.
Bagdouri and Oard~\cite{bagdouri-cikm15}
used Logistic Regression
on a dataset of 80 million Arabic tweets
from 91 thousand users,
out of which 3.64\% were deleted.
They used bag of words and tweet attribute features,
and user features to obtain a moderate F1-score of 0.45.


\Replacement{One of the first large scale studies of deleted tweets
was presented by Almuhimedi et al.~\cite{almuhimedi-cscw13}.
Over a period of one week,
Almuhimedi et al.~\cite{almuhimedi-cscw13}
collected all tweets posted by
292 thousand Twitter users,
which also included tweets later deleted by their authors.
On comparing the deleted and non-deleted tweets in their dataset,
the authors were able to find differences along various dimensions
such as location of origin of the tweet
and the Twitter client used for posting the tweet.
However, the authors also noted that the major cause
of tweet deletions in their dataset was
superficial reasons such as typos and rephrasings.}

In a very recent work, Zhou et al.~\cite{zhou-www16}
built a classifier for \emph{content-identifiable regrettable tweets}.
They defined content-identifiable regrettable tweets,
as those which third party human annotators think are regrettable.
\Replacement{The authors crafted a set of 10 closed vocabulary features
representing words related to sensitive topics
such as alcohol use, drug use, violence etc.}
For the classification task
they limited themselves to choosing
a set of 10,000 deleted and 10,000 non-deleted tweets
such that they had words related to the above 10 categories.
Using a decision tree classifier
they were able to achieve a F1-score of 0.714.

~\\
\noindent {\bf Novelty of present work:}
The present work also tries to build
a classifier to predict if a tweet will
be deleted or not.
However, unlike Zhou et al.~\cite{zhou-www16}
we do not restrict ourselves to the 18\% subset of deleted tweets
that they define as content-identifiable regrettable tweets.
Moreover, unlike Petrovic et al.~\cite{petrovic-unpub13}
Bagdouri and Oard~\cite{bagdouri-cikm15}, and
Zhou et al.~\cite{zhou-www16}
we use a mixed model with open-text bag of words
features along with closed-text features,
and author features to obtain a significantly higher classification
accuracy of 0.78 F1-score,
for all deleted tweets.
Further, we also look at the possibility of using
features from responses to tweets,
which boosts our accuracy to a F1-score of 0.817.

Additionally, the current work makes a number of novel contributions
related to understanding of deleted tweets and their authors.
First, one of the major focuses of this work
is to understand the underlying personality differences
between users who delete their tweets and those who do not.
Second, this work presents a thorough linguistic analysis
of deleted tweets and non-deleted tweets
from the same user set.
Third, we look at responses to deleted and non-deleted tweets
which provide additional strong signals about
the likelihood of a tweet being deleted.
Almuhimedi et al.~\cite{almuhimedi-cscw13} reported
to being unable to find differences in
``distributions of stereotypically regrettable topics''
due to the ``sheer volume and variety of reasons'' for which
tweets are deleted.
This work tries to solve this problem
by undertaking an extensive cleanup process
to remove spam, malware, automated tweets and superficial
deletions from our dataset.
This cleanup procedure allows us to bring forward the
characteristic qualities of regrettable content
present in deleted tweets.

~\\
Finally, a preliminary version of the current work
has been published as a short paper~\cite{bhattacharya-icwsm16}.
The current work significantly improves and extends the analysis
presented in~\cite{bhattacharya-icwsm16}.
For instance, while~\cite{bhattacharya-icwsm16} simply noted
the difference in language use between deleted and non-deleted tweets,
the current submission analyses the phenomenon at a deeper level,
by analyzing it both at an aggregate level,
using Normalized Tweet Difference (NTD),
as well as at an individual user level,
using Normalized User Difference (NUD).
Further, the current submission includes a completely new user survey
based analysis of deleted tweets
that tries to categorize them with respect to known categories
of regrettable communications established
in earlier works~\cite{sleeper-chi13}.
Additionally, the current study also includes
the design of a two stage classifier,
which uses textual, contextual, as well as user features
to predict whether a tweet is likely to be deleted in the future.

\section{Dataset}
\label{sec:dataset}

One of the major objectives of this work
is to understand tweet deletion behavior
of \emph{real and active} Twitter users.
To select a representative sample
of real and active Twitter users,
we began by selecting a random set of 2.5 million Twitter users
whose tweets were included in the
1\% Twitter Random Sample%
\footnote{\url{https://blog.gnip.com/tag/spritzer/}}
during the month of October 2014
and whose majority tweets were in English.
To filter out malicious users,
that is users who had been suspended or
users who had deleted their account,
their profiles were re-crawled during March 2015.
Further, we crawled up to 3,200 tweets for each user
to check if they had posted any \emph{unsafe} links.
For every url that is tweeted,
Twitter uses its own url shortening service t.co
to shorten it.
Apart from reducing the size of url posted,
t.co  also filters out and warns users
from visiting malicious urls%
\footnote{\url{https://support.twitter.com/articles/90491-unsafe-links-on-twitter}}.
We considered users to be potentially malicious
if they had posted any url
that was marked unsafe by Twitter.

Of the users who still had an active account
and were not marked potentially malicious,
we randomly selected 250,000 users
who also met the following criteria:
they had posted at-least 10 tweets in their lifetime
and had at least 10 followers and 10 followees.
These criteria were enforced to ensure that
only active Twitter users were selected.
\Replacement{During the four week period
of August 3, 2015 to August 30, 2015,
we followed these users and collected
all tweets posted by them
as well as replies and retweets of their tweets,
using the Twitter Streaming API.}


\Replacement{The Twitter Streaming APIs send out tweets in near realtime,
i.e. as soon as they are posted.
Thus, when tweets are later deleted by their authors,
the steaming api endpoints have to send out status deletion notices
to inform whoever is collecting the tweets
of the tweet deletion
\footnote{\url{https://dev.twitter.com/streaming/overview/messages-types\#status_deletion_notices_delete}}.
As there can be significant delays between when a tweet is posted
and when it is deleted,
we collected status deletion notices for an additional week,
that is during the five week period
of August 3, 2015 to September 6, 2015.}


The first part of Table~\ref{tab:dataset-stats} shows the total count of tweets
that were captured in our dataset.
Out of the selected 250 thousand users,
214 thousand users had posted a total of 43 million tweets,
during the four week period.
About 71.14\% users in our dataset had deleted at-least one tweet.
The total fraction of tweets that were deleted was 11.11\%.%

\begin{table}
\centering
\small

\begin{tabu} to \linewidth{@{}lr@{}}
\toprule

\textbf{Dataset before cleanup} \\
\# Tweets posted                      & 43,119,286 \\
\# Tweets deleted                     & 4,793,101 (11.11\%) \\
\# Users who posted at-least 1 tweet  & 214,471 \\
\# Users who deleted at-least 1 tweet & 152,591 (71.14\%) \\

\midrule

\textbf{Non-english tweets} \\
\# Tweets posted                      & 8,891,985 \\
\# Tweets deleted                     & 908,659 (10.21\%) \\
\# Users who posted at-least 1 tweet  & 180,997 \\
\# Users who deleted at-least 1 tweet & 84,495 (46.68\%) \\

\midrule

\textbf{Automated tweets} \\

\# Tweets posted                      & 2,330,486 \\
\# Tweets deleted                     & 310,585 (13.32\%)\\
\# Users who posted at-least 1 tweet  & 49,144 \\
\# Users who deleted at-least 1 tweet & 12,041 (24.50\%) \\

\midrule

\textbf{Retweets} \\

\# Tweets posted                      & 18,939,729 \\
\# Tweets deleted                     & 2,898,029 (15.30\%) \\
\# Users who posted at-least 1 tweet  & 187,916 \\
\# Users who deleted at-least 1 tweet & 124,953 (66.49\%) \\

\midrule

\textbf{Superficial deletions} \\

\# Tweets posted                      & 288,934 \\
\# Users who posted at-least 1 tweet  & 69,976 \\

\midrule

\textbf{Dataset after cleanup} \\

\# Tweets posted & 17,147,771 \\
\# Tweets deleted & 1,210,434 (7.05\%) \\
\# Users who posted at-least 1 tweet & 194,495 \\
\# Users who deleted at-least 1 tweet & 91,785 (47.19\%) \\

\bottomrule
\end{tabu}
\caption{Total number of tweets posted and deleted by users in our dataset,
  before and after,
  removal of non-english tweets, automated tweets, retweets, and superficial deletions.}
\label{tab:dataset-stats}
\end{table}



\subsection{Dataset cleanup}

\noindent As stated earlier, the objective of this study is to 
analyse the {\it human factors} leading to deletion of tweets.
In other words, we wanted to focus on those tweets which were deleted
because of human factors like regrettable content. 
Since tweets can be deleted due to various other reasons, 
we undertook a systematic cleanup procedure of removing
certain specific types of tweets, as described below. 

\emph{Non-English tweets:}
For this work, we focus only on English tweets.
Tweet objects, obtained from Twitter streaming and REST APIs,
have language fields populated by Twitter's
automatic language detection methods~\cite{tweet-lang-field}.
To filter out non-English tweets,
we removed any tweet that did not
have `en' as the language attribute.

\emph{Automated tweets:}
As Twitter has grown in popularity,
developers have created automated tweeting systems
to use Twitter in creative new ways.
However, for this analysis,
we removed automated tweets from our dataset, since
automated tweets are usually posted for
promotion purposes and hence, are unlikely to be
deleted due to human factors (on which we want to focus).
The key insight that used to filter automated tweets is ---
most automated tweets
are posted using specialized Twitter clients.
For this work, we whitelisted Twitter clients
which are popularly used and whose main functionality is
to allow their users to post tweets.
These include the official Twitter clients for
web and mobile platforms, as well as other
popular third party Twitter clients such as
TweetDeck, HootSuite, etc.
The top Twitter sources that were excluded
from the whitelist included:
(i)~automated tweeting systems like:
RoundTeam, If this then that, Buffer, and twittbot.net
(ii)~account management tools such as: fllwrs and Crowdfire App,
and (iii)~clients for specific websites like:
Twittascope, Ask.fm, and WordPress.com.

\emph{Deleted retweets:}
When a tweet is deleted,
any retweet of the said tweet is also deleted by Twitter%
\footnote{\url{https://support.twitter.com/articles/18906-deleting-a-tweet}}.
\Replacement{Since, status deletion notices sent out by Twitter
do not contain the information about
whether the original tweet or the retweet was deleted,
analysis of deleted retweets is left out from this work.}

\emph{Superficial deletions:}
\Replacement{Twitter doesn't provide a method to edit tweets.
Users thus have to delete the bad tweet and create a new one,
to fix typos or grammatical errors.
Superficial deletions like these are ignored by the current work
as we focus mainly on analyzing and understanding regrettable communications.}
To check if a tweet deletion is superficial or not
we take an approach similar to Almuhimedi et al.~\shortcite{almuhimedi-cscw13}.
For every deleted tweet, we check the next (chronologically) three
tweets made by the same user who deleted the tweet.
If we find any of the three tweets to be similar to the deleted tweet,
we term the deletion superficial.
We call two tweets to be similar if the edit distance
between texts of the tweets is less than 5 or the cosine similarity
between terms of the tweets is greater than 0.6.
Similar to Almuhimedi et al.~\shortcite{almuhimedi-cscw13},
we found that according to the above measure,
14.45\% of tweet deletions in our dataset
were classified as superficial errors or typos.

~\\
The second part of Table~\ref{tab:dataset-stats} shows the final count of
tweets and users in our dataset
after removal of non-English tweets, automated tweets, retweets,
and superficial deletions.

Finally, to understand the impact of bot accounts in our analysis,
we randomly sampled a set of 100 users from our clean dataset,
and manually checked them for bot accounts.
We found that out of the 100 users accounts 97 were personal accounts
belonging to normal users, while 3 belonged to different organizations.
However, none of the 100 accounts were bot accounts.
This confirms the effectiveness of our cleanup methodology
and ensures that the inferences obtained from our analysis
are unlikely to be significantly effected by bot accounts.

\section{Understanding causes of tweet deletion}
\label{sec:delcategorize}

Earlier studies \cite{sleeper-chi13,knapp-jc86},
in an effort to better understand regrettable communications,
tried to categorize them
into a finite set of classes.
In the current section,
we try to answer the question:
\emph{what fraction of deleted tweets
relate to these categories of regrettable communications?}

\subsection{Categories of regrettable communications}

Sleeper et al.~\shortcite{sleeper-chi13},
utilized the categories of regrettable communications
presented in \cite{knapp-jc86},
to classify regrettable communications
in Twitter and in the offline world.
The classes of regrets used were as follows:
\begin{itemize}[noitemsep]
  \item \emph{Blunder}: errors in formulating statements,
    including typos, and factual errors.
  \item \emph{Direct attack}: critical statements directed
    towards a person.
  \item \emph{Group reference}: critical statements directed
    towards a group (ethnicity, race, etc.)
  \item \emph{Direct criticism}: criticism
    directed towards ``something specific about a person''.
  \item \emph{Reveal/explain too much}: revealing personal,
    potentially embarrassing information.
  \item \emph{Agreement changed}: statements of agreement
    that were later changed by the author.
  \item \emph{Expressive/catharsis}: emotional statements
    not directed towards any specific person.
  \item \emph{Lie}: intentionally stating something
    that is not true.
  \item \emph{Implied criticism}: statements of criticism
    that are implicit, such as teasing remarks.
  \item \emph{Behavioral edict}: asking someone to behave in
    a particular manner.
\end{itemize}

In the study conducted in \cite{sleeper-chi13},
the authors asked survey participants
to recall and report a regrettable communication
that they had made on Twitter
or in the offline world.
A limitation of this method is that
participants are more likely to report
memorable regrets rather than the most prevalent regrets.


\subsection{Categorizing deleted tweets}

To understand the relative prevalence
of different types of regrettable communications
in tweets,
we manually annotated tweets with the
categories discussed above.
However, for the annotation process,
we merged the groups ``Direct attack'', ``Direct criticism'',
and ``Implied Criticism''.
This was primarily done because in our preliminary studies
we found that the annotators were having problems
distinguishing between these three categories,
from the limited context available from 140 character tweets.
Also removed was the ``Blunder'' category,
as we had identified and removed superficial deletions
during our data cleanup process.
Further, as it is not possible for a third party
annotator to understand, without context,
if a person has lied or have changed their viewpoints,
we removed those categories as well.
Every tweet was independently annotated by
three annotators.
The annotators were graduate students
from the authors' institute, but none of them were authors of this paper.
For each tweet evaluated by the annotators,
they were asked a yes or no question
on whether the tweet belonged to a given category.
The annotators had three options, yes, no, and can't say.
Additionally, the annotators were asked to answer if
they thought ``the author posting the given tweet may regret
posting it later''.

Overall one hundred deleted and one hundred non-deleted tweets,
were annotated.
The annotators had unanimous agreement on 56\% cases
and majority agreement on 96\% cases.
We consider a tweet to belong to a category
if at-least two of the three annotators thought
that it did belong to the category.

\begin{table}
\centering
\begin{tabu} to \linewidth{@{}lrr@{}}
\toprule
\rowfont[c]{}
& Del Tweets & Non-Del Tweets \\

\midrule

Reveal too much    & 30 & 27 \\
Expressive         & 19 & 32 \\
Direct attack      & 11 & 9 \\
Behavior edict     & 2  & 8 \\

\midrule

Regrettable        & 16 & 6 \\

\bottomrule
\end{tabu}
\caption{Number of tweets
  (out of a random sample of 100)
  which fall under a given
  category according to majority decesion
  of the annotators classifying the tweet.
  Tweets for which there was no majority decision are not counted in the table.}
\label{tab:tweet-categorize}
\end{table}






\begin{table}[t]
\small
\begin{tabu} to \linewidth {X[l]}
\toprule
\textbf{Reveal too much:}\\
\textbullet~@[username] ....it was aliens wasnt it.
(Im at work and wont get home till 6  friiiiick me)\\
\textbullet~Help me save our lands from the Koch Brothers!
via @[username] [url]\\
\textbf{Expressive:}\\
\textbullet~@[username] im done fuckin with you!\\
\textbullet~School isn't for me. I hate it. It's annoying. It's boring.
It's too easy. It's a waste of my time.\\
\textbf{Direct attack:}\\
\textbullet~Not only does @[username] buy useless shit for herself
she also buys it for her cats  :‑J ;)\\
\textbullet~@[username] hey faggot!!  Lol :D [url]\\
\textbf{Behavior edict:}\\
\textbullet~A hopeless person sees difficulties in every chance,
but a hopeful person sees chances in every difficulty. @[username]\\
\textbullet~``Better to have little, with godliness,
than to be rich and dishonest.'' Proverbs 16:8~NLT @[username]\\
\bottomrule
\end{tabu}
\caption{Sample deleted tweets belonging to the different regrettable catergories.
The tweets have been anonymized by replacing mentioned usernames and urls}.
\label{tab:categorize-tweet-sample}
\end{table}





Table~\ref{tab:tweet-categorize}
and Table~\ref{tab:categorize-tweet-sample} show the results of
the annotation experiment
and some sample deleted tweets from the different categories.
Interestingly we observe that,
while deleted tweets had slightly more tweets
in the categories ``Reveal too much'' and ``Direct attack'',
 non-deleted tweets had a larger fraction
belonging to categories ``Expressive'' and ``Behavior edict''.

Surprisingly, we found that for only 16 out of the 100 deleted tweets
the majority annotators agreed that
the author posting the tweet might regret it later.
Additionally, for 6 out of 100 non-deleted tweets
the majority annotators agreed that those tweets
might cause their authors to regret posting them later.
While this difference in statistically significant
(using Fisher's Exact test with $\text{Odds Ratio}=0.33$ and $p=0.04$)
it seems that for the majority of deleted tweets (84\%)
the personal nature of the underlying regrettable cause,
which triggered the tweet deletion,
is not understandable by any third party viewers,
from only the content of the tweets.
Further, as discussed earlier in this section,
for cases belonging to regrettable categories such
as ``Lie'' and ``Agreement Changed'',
it is nearly impossible for third party annotators
to identify the underlying regrettable cause without sufficient context,
and were thus not even included in our annotation experiment.
This result is in agreement with a similar finding
made by Zhou et al.~\shortcite{zhou-www16}.
Due to the difficulty in differentiating between
deleted tweets that were deleted due to regrets,
and those that were deleted for other causes,
Zhou et al.~\shortcite{zhou-www16} studied a subset
of deleted tweets that they defined as
\emph{content-identifiable regrettable tweets}.
Zhou et al. defined content-identifiable regrettable tweets
as those which third party human annotators think are regrettable.
They noted that only 18\% deleted tweets in their corpus
matched the above definition.

\section{Characterizing users who delete their tweets and those who do not}
\label{sec:delusers}

While we noted in the previous section,
that it may be significantly difficult for third party humans
to figure out the regrettable cause behind the
deletion of particular tweets,
from the text of the tweet alone,
it may still be possible to extract high level
of characteristics of tweets that are deleted
and their authors who deleted them.
In this section, we answer the question ---
\emph{does there exist any characterizing differences
between users who delete tweets and those who do not?}
To answer this question,
we analyze tweets
from two subsets of users from our dataset.
(i)~\emph{non-deleters}: the set of 102 thousand users
who had posted at least one tweet,
but had either not deleted any tweet or all their tweet deletions (if any)
were classified as superficial deletions,
and (ii)~\emph{deleters}: the set of 92 thousand users
who had deleted at least one tweet,
that was not a superficial deletion.

\begin{figure}
  \centering
  \subfloat[Follower distribution\label{fig:user-follower-dist}]{%
   \includegraphics[width=0.35\textwidth]{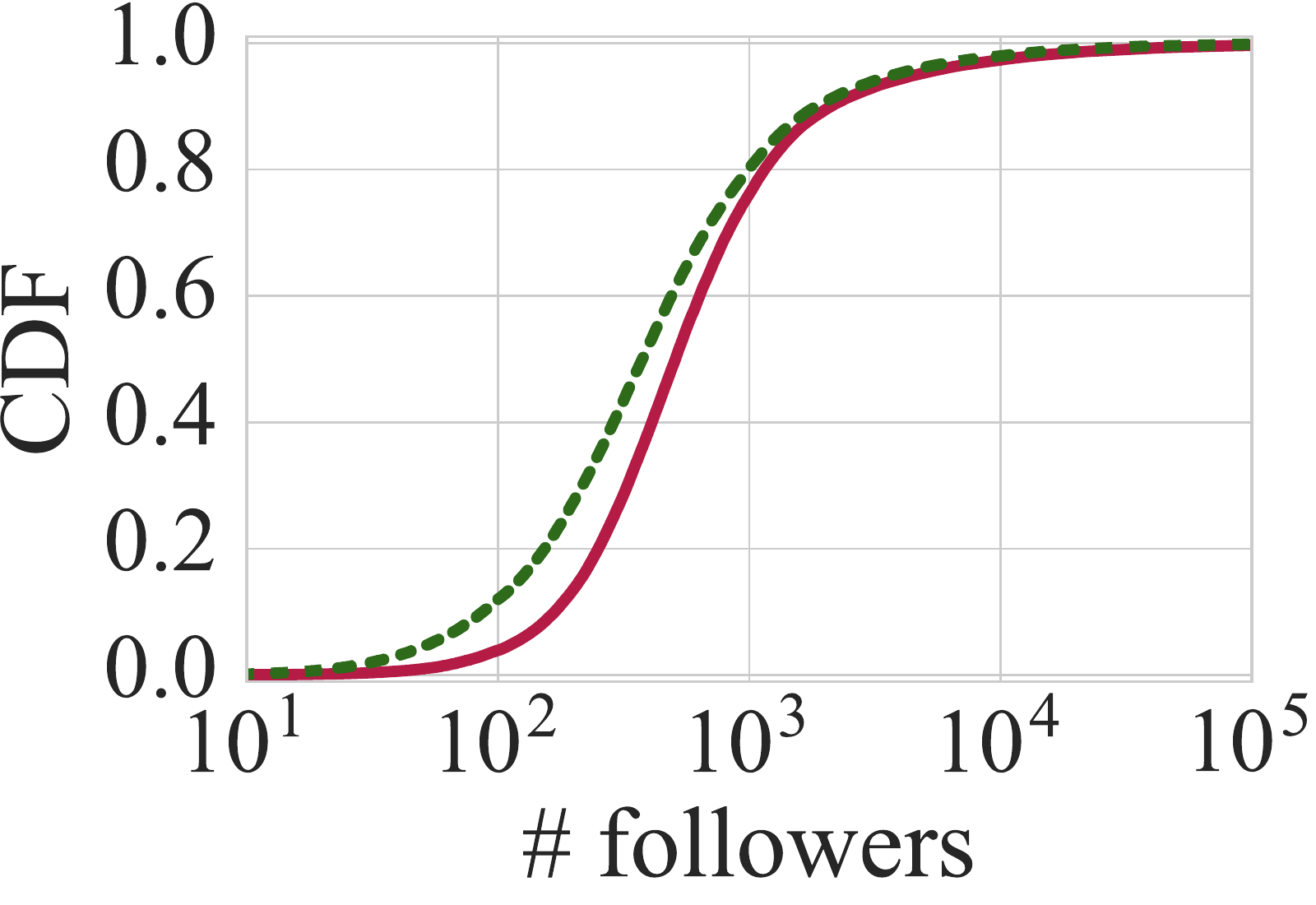}
  }
  ~~
  \subfloat[Followee distribution \label{fig:user-followee-dist}]{%
    \includegraphics[width=0.35\textwidth]{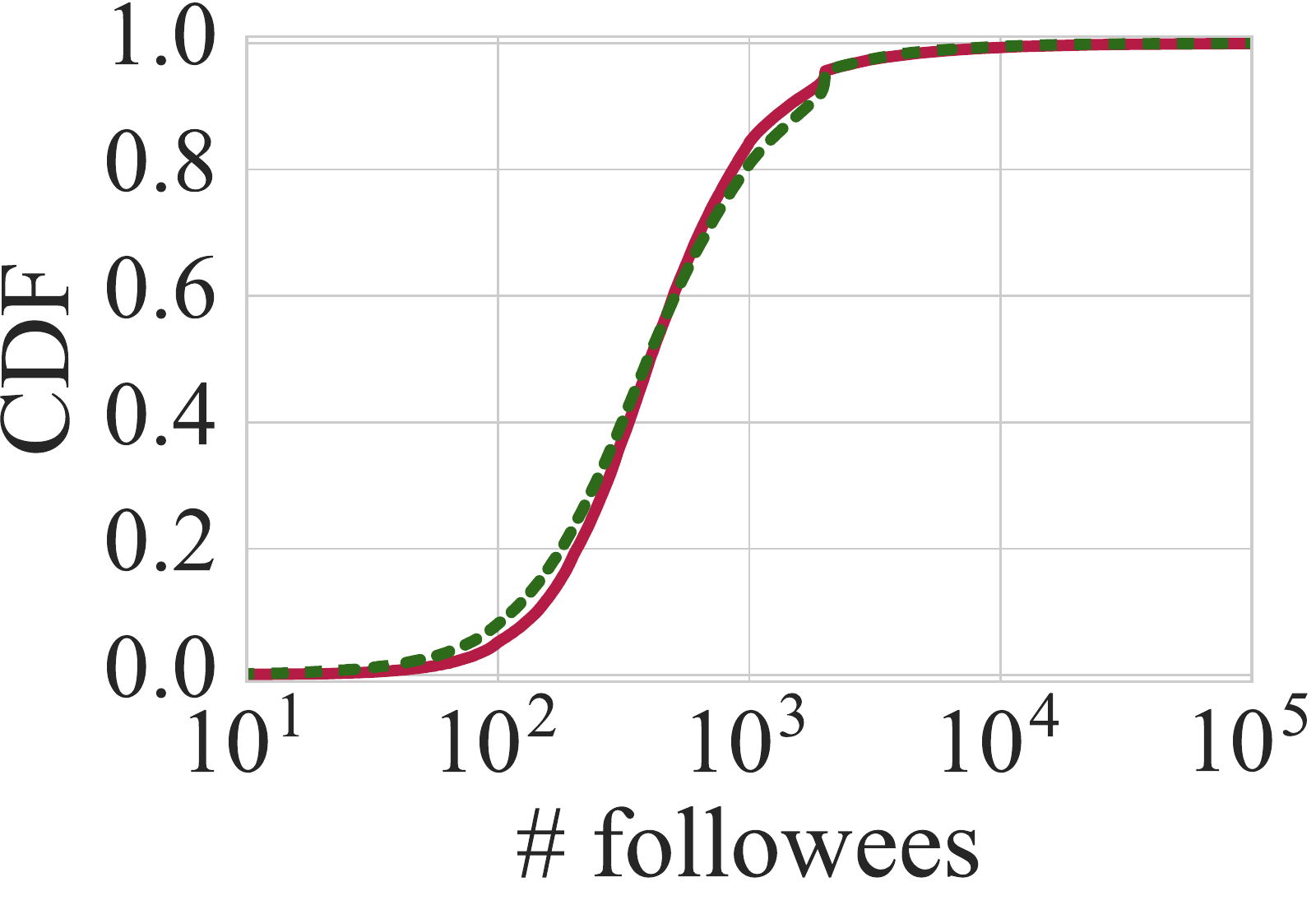}
  }

  \subfloat[Listed count distribution \label{fig:user-listed-count-dist}]{%
    \includegraphics[width=0.35\textwidth]{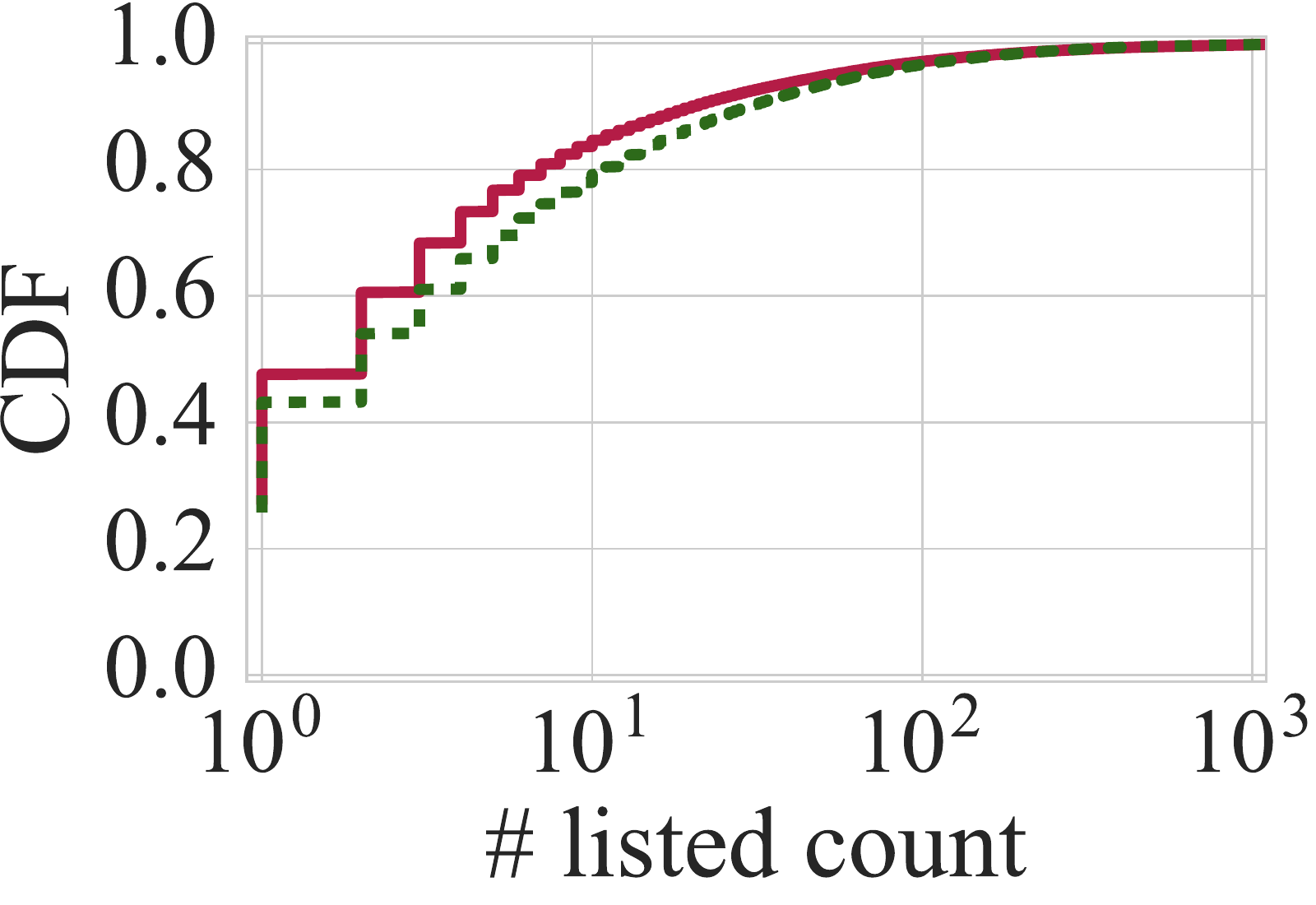}
  }
  ~~
  \subfloat[Tweet rate distribution \label{fig:user-tweet-rate-dist}]{%
    \includegraphics[width=0.35\textwidth]{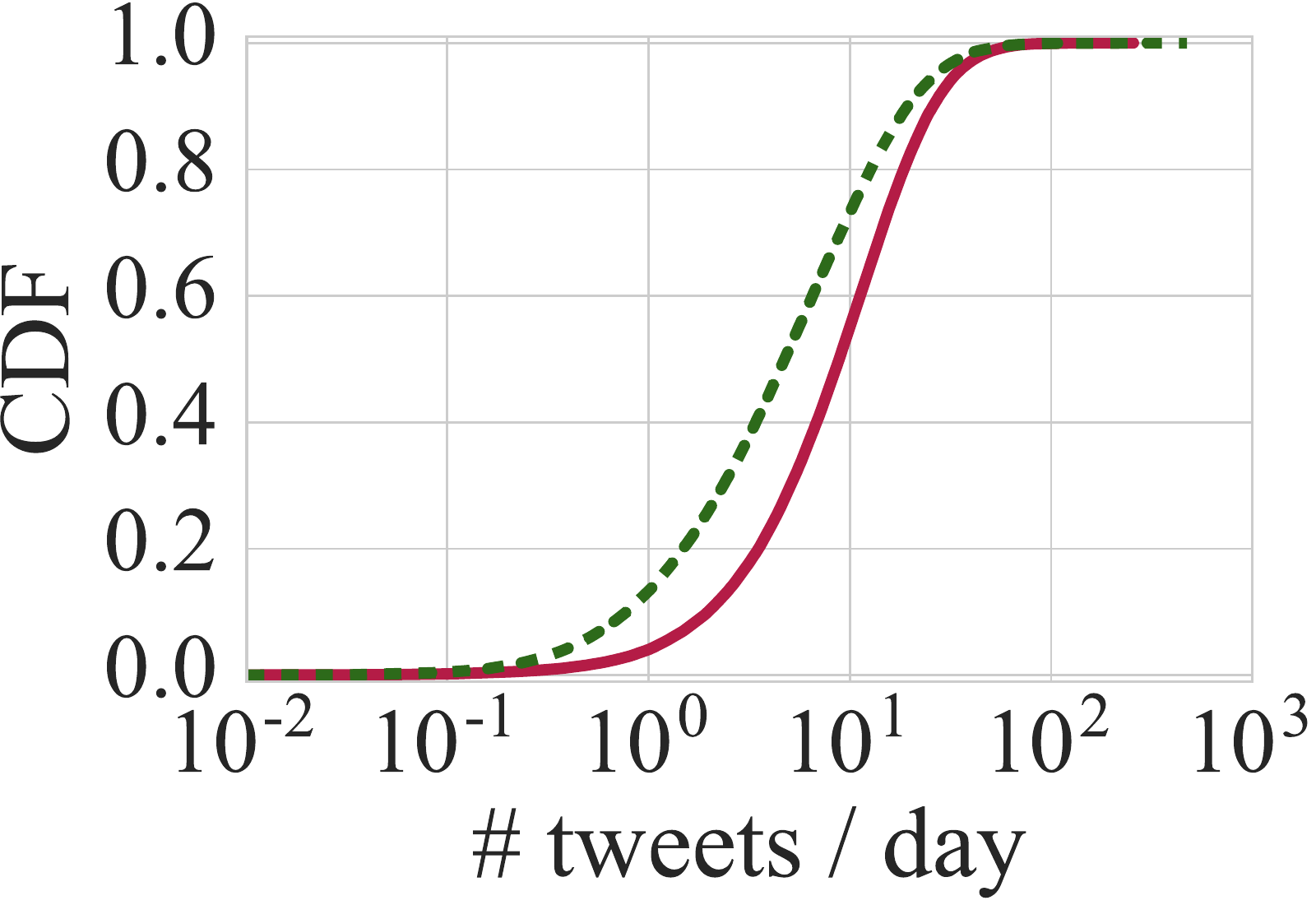}
  }

  \caption{Distribution of user attributes for users in the deleter
           and non-deleter sets.
           All four distribution pairs presented here
           have statistically significant differences
           when compared using Mann-Whitney U test with $p<0.001$.}
  \label{fig:user-attr-dist}
\end{figure}

\subsection{Big-Five personality traits}
It is expected, that any significant differences in tweet deletion practices
among the user groups (if they exist)
would stem from their underlying personality differences.
The Big-Five personality traits, also known as the Five factor model,
is a very popular system for modeling human personality
in terms of five orthogonal dimensions~\cite{big-five}.
The five characteristic traits,
using which the Big-Five system models human personality, are:
\begin{itemize}
  \item \emph{Openness to experience:} Users scoring high on openness
    generally are appreciative of art and are open to new experiences
    and diverse ideas.
  \item \emph{Conscientiousness}: Conscientious individuals have
    high degrees of self-discipline while tending to be high achievers
    and meticulous planners.
  \item \emph{Exterversion:} Extroverted people are highly social,
    and are generally friendly, drawing energy from social situations.
  \item \emph{Agreeableness:} Agreeable users have an optimistic
    outlook and try to maintain social harmony.
  \item \emph{Neuroticism:} Neurotic individuals tend to be moody
    with high propensity to feel negative emotions.
\end{itemize}

\Replacement{In the Big-Five model, the five traits
capture five orthogonal dimensions of human personality.}
The Big-Five model has been well accepted
for being able to account for
human behavior in a number of studies~\cite{big-five}.
Traditionally, to elicit a person's relative score
in terms of the five dimensions,
the user has to fill out a personality questionnaire
consisting of self descriptive questions~\cite{big-five}.
However, recently a number of studies have tried to characterize
the Big-Five personality traits of social media users,
by developing classification models
that correlate the user's social media profile features
with their personality traits%
~\cite{ryan-chb11,gosling-cbsn11,schwartz-pone13,quercia-passat11}.
\Replacement{The rest of the current section,
tries to uncover the differences in personalty
between Twitter users in the deleter and non-deleter user sets
by comparing their social and linguistic attributes.}

\subsection{Differences in social characteristics}

\Replacement{Earlier studies~\cite{quercia-passat11} have noted
the strong relations between a Twitter user's social features
and their Big-Five personality traits.
Quercia et al.~\shortcite{quercia-passat11} in their effort
to predict the personality of Twitter users,
found strong correlations between a users Big-Five personalty traits
and their social features (number of followers, number of followees, etc.).
Here, we leverage the insights of~\cite{quercia-passat11}
by comparing the differences
between users of the deleter and non-deleter user sets
in terms of their social features.}

Figure~\ref{fig:user-attr-dist} shows the distribution
of these attributes between the two user sets.
We find that, users in the deleter set
have significantly higher follower count (Figure~\ref{fig:user-follower-dist}),
with median follower counts
for users in the deleter and non-deleter sets
being 508 and 375 respectively.
Also, when considering the full distribution,
we see in Figure~\ref{fig:user-listed-count-dist}
that users in the deleter set
have significantly lower listed count
(the number of Twitter Lists they are a member of).
Figure~\ref{fig:user-tweet-rate-dist} shows that
the median tweet rate of users in the deleter set is nearly
twice the tweeting rate of the users in the non-deleter set
(8.85 tweets/day compared to 4.74 tweets/day).
However, the difference between the distributions
of followee count (Figure~\ref{fig:user-followee-dist}) is not that prominent;
we observe that median followee count for the users
in the deleter and non-deleter sets
are 403 and 394 respectively.


\Replacement{The statistically significant differences
presented above, in conjunction with the 
strong and significant correlations
presented by Quercia et~al.~\shortcite{quercia-passat11},
suggests that users in the deleter set are more likely to be
\emph{extroverted}, \emph{neurotic}, and less \emph{open}
compared to users in the non-deleter set.}

\begin{table}
\centering
\footnotesize

\begin{minipage}{0.63\textwidth}
\begin{tabu} to \linewidth{@{}lX[r]X[r]X[1.5c]@{}}
\toprule

\rowfont[c]{}

& Non-Deleters & Deleters & Predicted Deleter \mbox{Personality}\\

\midrule
\multicolumn{4}{l}{\textbf{LIWC Features}} \\

2nd Person Pronouns    & 1.39\% & 1.92\% & $A$, $C$ \\
Articles               & 3.43\% & 3.29\% & - \\
\mbox{Auxiliary Verbs} & 5.87\% & 7.40\% & $\bar{C}$ \\
Future Tense           & 0.49\% & 0.69\% & $\bar{C}$ \\
Negations              & 1.11\% & 1.71\% & $\bar{C}$ \\
Quantifiers            & 1.45\% & 1.68\% & $O$ \\

\midrule

\mbox{Social Process} & 6.42\% & 7.42\% & $E$ \\
Family                & 0.02\% & 0.25\% & $E$ \\
Humans                & 0.50\% & 0.73\% & $O$ \\

\midrule

Negative Emotions & 1.41\% & 2.17\% & $\bar{C}$ \\
Sadness           & 0.15\% & 0.36\% & $\bar{C}$ \\

\midrule

Cognitive Mechanisms & 9.28\% & 10.49\% & $\bar{C}$ \\
Causation            & 0.83\% & 1.01\%  & $\bar{A}$, $O$ \\
Discrepancy          & 0.85\% & 1.25\%  & $\bar{C}$ \\
Certainity           & 0.82\% & 1.10\%  & $O$ \\

\midrule

Hearing & 0.23\% & 0.41\% & $N$ \\
Feeling & 0.25\% & 0.43\% & $\bar{C}$, $N$ \\

\midrule

Biological Processes & 1.49\% & 2.19\% & $\bar{O}$ \\
Body                 & 0.35\% & 0.74\% & $\bar{O}$ \\
Health               & 0.19\% & 0.35\% & $\bar{E}$ \\
Ingestion            & 0.08\% & 0.26\% & $A$ \\

\midrule

Work        & 0.95\% & 0.91\% & $\bar{C}$, $\bar{O}$ \\
Achievement & 1.08\% & 1.02\% & $A$ \\
Money       & 0.28\% & 0.33\% & $\bar{A}$ \\
Religion    & 0.06\% & 0.21\% & $N$ \\

\midrule
\multicolumn{4}{l}{\textbf{Non LIWC Features}} \\
\mbox{w/ +ve Sentiment} & 69.23\% & 63.63\% & $\bar{O}$ \\
\mbox{w/ -ve Sentiment} & 29.16\% & 36.36\% & \\
\mbox{w/ Hashtags}      & 5.0\%   & 2.47\%  & $O$ \\
w/ Urls                 & 29.05\% & 16.66\% & $\bar{C}$ \\

\bottomrule
\end{tabu}
\end{minipage}
\hfill
\begin{minipage}[c]{0.34\textwidth}
\caption[]{\footnotesize Median percentage of words
  from different LIWC categories
  tweeted by users in the non-deleter and deleter sets.
  Also presented are the median percentage of tweets
  with positive and negative sentiments,
  as well as median percentage of tweets with hashtags and urls.
  The final column presents the likely personality trait
  predicted for the deleter set
  due to the difference in the catergory.
  
  \vspace{0.3\baselineskip}
  For example, $A$ means higher likelihood of users in the deleter set
  having agreeableness trait,
  while $\bar{C}$ represents lower likelihood of users in the deleter set
  having conscientious trait.
  Here $O$ stands for openness,
  $C$ for conscientiousness,
  $E$ for extraversion,
  $A$ for agreeableness,
  and $N$ for neuroticism.

  \vspace{0.3\baselineskip}
  $\bar{C}$ is predicted by 10 features:
  auxiliary verbs, future tense, negations, negative emotions,
  sadness, cognitive mechanisms, discrepancy, feeling, work,
  and presence of urls,
  while the converse is predicted by only one feature:
  second person pronouns.
  $N$ is predicted by 3 features: hearing, feeling, and religion.
  The results for other personality traits are mixed.

  \vspace{0.3\baselineskip}
  All pairwise differences
  between deleter and non-deleter sets
  presented here are statistically significant,
  when compared using Mann-Whitney U test
  with $p < 0.001$.}
  \label{tab:user-liwc}
\end{minipage}
\end{table}

\subsection{Differences in linguistic style}
\Replacement{The correlations in writing style and
Big-Five personalty scores of users have been well studied
in social media~\cite{golbeck-passat11,schwartz-pone13}.
Golbeck et al.~\shortcite{golbeck-passat11}
demonstrated that similar correlations exist between
the Big-Five personality traits
of Twitter users and their tweets
when analyzed using the LIWC toolkit%
\footnote{\url{http://www.liwc.net}}.}
Here we try to use the significant correlations
presented by Golbeck et al.~\shortcite{golbeck-passat11}
to gain further insights into
the differences between the users in the deleter and non-deleter sets.


Table~\ref{tab:user-liwc} shows the differences
in the median percentage of words
for different LIWC
categories, used by users in the deleter and non-deleter sets.
Also presented are the median percentage of tweets
with positive and negative sentiments,
as well as percentage of tweets with hashtags and urls.
To understand the differences in the sentiment of tweets
posted by the users,
we used the Vader sentiment analysis tool~\cite{hutto-icwsm14}.
Only those categories for which
significant correlations with personality types
were shown by Golbeck et al.~\shortcite{golbeck-passat11} are presented.
The last column of Table~\ref{tab:user-liwc} shows the
personality trait predicted for the deleter user set
when using the significant correlations
presented by Golbeck et al.~\shortcite{golbeck-passat11}
for the given category.
For example $A$ represents higher agreeableness
while $\bar{C}$ represents less conscientiousness.
It can be observed that while the differences are small,
significant differences exist across all the categories.

From the differences presented in Table~\ref{tab:user-liwc}
we observe that the users in
the deleter set are likely to be \emph{less conscientious}
as predicted by the ten features
(while the converse is predicted by only one feature)
Also, users in the deleter set are more likely to be \emph{more neurotic}
as predicted by the three features.
However, the predictions from the other three personality traits
are not clear from the results in Table~\ref{tab:user-liwc}
as they contain mixed signals.


\subsection{Summary}
Overall, our analysis shows that
users who delete tweets are more likely to be
extroverted, neurotic, and less conscientious.
We hypothesize that,
extroverted and spontaneous (less conscientious) users are
more likely to speak up their minds,
altering their opinions
as more information becomes available to them.
Further, as neurotic users are more likely to be prone to stress and worrying,
it is intuitive that they might post more critical/offensive remarks
in their moments of stress.
Sleeper et al.~\shortcite{sleeper-chi13} noted that
both change of opinions and posting of critical/offensive content
can become regrettable for the users posting them.
One of the major actions taken by users when they later realize them,
is to delete their tweets.
This falls in line with our findings of the personality of users
who delete their tweets.

One limitation of the analysis in this section is that
we do not have ground truth personality information
of the users in our dataset.
We use strong and significant correlations reported in earlier
studies~\cite{quercia-passat11,golbeck-passat11}
to gain understanding of user personality.
However, we feel that although our observations
about user personalities are indirect,
they provide valuable and usable insights into the issue.

\section{Comparing tweets which are deleted to those that are not}
\label{sec:delcontent}

In this section,
we try to answer the question:
\emph{does there exist any characteristic
difference between tweets that are deleted
and those that are not?}
To answer this question,
we compared 1.2 million deleted
and 15.9 million non-deleted tweets
posted \emph{only by the users in the deleter set},
across different dimensions,
such as: presence of mentions, lexical density,
use of different linguistic features, etc.

These differences are compared at two levels.
To understand the overall difference
with respect to a given attribute,
we first compare the relative presence of the attribute
in deleted and non-deleted tweets at an aggregate level.
For this purpose, we define \emph{Normalized tweet difference (NTD)}
of deleted tweets with respect to non-deleted tweets.
For a given attribute $a$,
let $\text{DelTweetFrac}_a$
be the fraction of deleted tweets with the given attribute,
and $\text{NonDelTweetFrac}_a$
be the fraction non-deleted tweets with the given attribute.
Then, we define $\text{NTD}_a$ as follows:
\[
\text{NTD}_a = \frac
               {\text{DelTweetFrac}_a - \text{NonDelTweetFrac}_a}
               {\text{NonDelTweetFrac}_a}
               \times 100
\]
For any attribute $a$, $\text{NTD}_a$ is in the range $[-100, \infty)$.
A positive $\text{NTD}_a$ indicates
that the attribute $a$ has higher relative prevalance
in the deleted tweet set,
compared to the non-deleted tweet set.
A negative $\text{NTD}_a$ represents the converse.

However, as tweet deletion is a very personal choice,
we also measure the differentiating power of different attributes
at an individual level using \emph{Normalized user difference (NUD)}.
For the given attribute $a$,
let $\text{DelUserFrac}_a$ be the fraction of users
for whom the attribute $a$
is found significantly more in their deleted tweets,
and $\text{NonDelUserFrac}_a$ be the fraction of users
for whom the attribute $a$
is found significantly more in their non-deleted tweets.
Then, we define $\text{NUD}_a$ as follows:
\[
\text{NUD}_a = \frac
               {\text{DelUserFrac}_a - \text{NonDelUserFrac}_a}
               {\text{NonDelUserFrac}_a}
               \times 100
\]
As with $\text{NTD}_a$, for any attribute $a$,
$\text{NUD}_a$ is in the range $[-100, \infty)$.

In general, for finding users with significant differences
in their deleted and non-deleted tweets,
for computation of $\text{NUD}_a$,
we use Fisher's Exact test with $\alpha=0.05$,
when the comparison is between ratio of items
with and without the given attribute
(such as in
Table~\ref{tab:good-nort-tweet-diff},
first six attributes of Table~\ref{tab:postag-cmp},
and Table~\ref{tab:liwc-cmp}).
However, when having to compare the difference in median of distributions
(such as lexical density of tweets
and percentage of dictionary words,
as in Table~\ref{tab:postag-cmp})
we use Mann-Whitney U test with $\alpha=0.05$.
Further, when looking for users with significant difference
in their deleted and non-deleted tweets,
we restricted our analysis to the set of 19,739 users
(out of 92 thousand users in the deleter set)
for whom we have at-least 10 deleted and 10 non-deleted tweets
in our dataset.

Tables~\ref{tab:good-nort-tweet-diff}, \ref{tab:postag-cmp}, and \ref{tab:liwc-cmp}
compare the normalized tweet difference and normalized user difference
for different attributes.
Other than the few cases marked with \xitem,
we find that in general both metrics agree with each other.
\todoNG{Why??}

\begin{table}
\centering
\small

\begin{tabu} to \linewidth{@{}lrr@{}c@{}}
\toprule

\rowfont[c]{}
& NTD
& NUD & \\

\midrule


Tweets w/ hashtags & -50.29\% &  -6.31\% & \\
Tweets w/ urls     & -51.32\% &  78.54\% & \xitem \\
Tweets w/ mentions & -27.90\% & -71.63\% & \\
Replies            & -24.79\% & -71.07\% & \\

\bottomrule
\end{tabu}
\caption[]{
Normalized Tweet Difference (NTD) and Normalized User Difference (NUD)
with respect to fraction of tweets that are replies
and fraction of tweets that contain hashtags, urls, and mentions.
Interestingly, we find that the trends obtained with respect to urls reverses
when seen from tweet (NTD) and user (NUD) point of view.
The fraction of deleted and non-deleted tweets
(used to compute NTD)
have statistically significant difference
with respect to all given attributes,
when compared using Fisher's Exact test with $p < 0.001$.}
\if{0}
  Also presented is the percentage of users (Del Signif Users)
  who use hashtags, urls, or mentions
  signficantly more in their deleted tweets
  or for whom their deleted tweets
  are significantly more likely to be replies,
  when compared to their non-deleted tweets.
  Similarly, the percentage of users (Non-Del Signif Users)
  for whom the reverse is true is also shown.\protect\footnotemark}
\fi
\label{tab:good-nort-tweet-diff}
\end{table}





\subsection{Comparing tweet attributes}


\Replacement{Due the short and succinct nature of tweets,
enforced by a 140 character limit,
Twitter users often use hashtags and urls
to supplement the information contained in their tweets.
Our previous studies~\shortcite{ghosh-cikm13,zafar-tweb15} had noted that
tweets containing hashtags and urls
are generally more informative.
Hashtags help in contextualizing tweets by
linking them to the bigger discussion,
while additional sources of information are often
provided by the embedded urls.
To analyze the differences in information content,
we compared deleted and non-deleted tweets
relative to these attributes.}

Table~\ref{tab:good-nort-tweet-diff} compares
deleted and non-deleted tweets
with respect to the percentage of them being
replies or those that contain mentions, hashtags and urls.
We find that, when compared to non-deleted tweets,
the percentage of deleted tweets that contains
hashtags and urls is nearly half
(50.29\% drop for hashtags and 51.32\% drop for urls).
This indicates that overall,
\emph{information content of tweets in the deleted set
is significantly lower}
when compared to tweets that are not deleted.
Interestingly, while at the aggregate level deleted tweets
contain significantly less urls,
for a larger fraction of users
their use of urls is significantly more in their deleted tweets.
On checking, we found that in many cases the urls pointed
to photos uploaded on Twitter, which were also deleted
along with the tweets.

One may postulate that this lack of informative content
can be attributed to the more conversational nature of deleted tweets.
Replying to tweets posted by others and
mentioning other's user names in tweets
are ways of starting and continuing conversations on Twitter.
To compare the percentage of conversational tweets that are deleted,
we compared the fraction of tweets
that contain mentions and those that are replies,
in the deleted and the non-deleted tweet sets.
Table~\ref{tab:good-nort-tweet-diff} shows that
fewer deleted tweets (24.79\% less) are replies
and a lesser fraction of them (27.9\% less) contain mentions.
Similar results are found in the user level comparison.
This indicates that overall \emph{conversational tweets
are less likely to be deleted}.

\if{0}
where we find that for a significantly larger fraction of users
(27.66\% for replies and 29.05\% for tweets with mentions)
their conversational tweets are less likely to be deleted.
\fi

\begin{figure}
\centering
\includegraphics[width=0.5\textwidth]{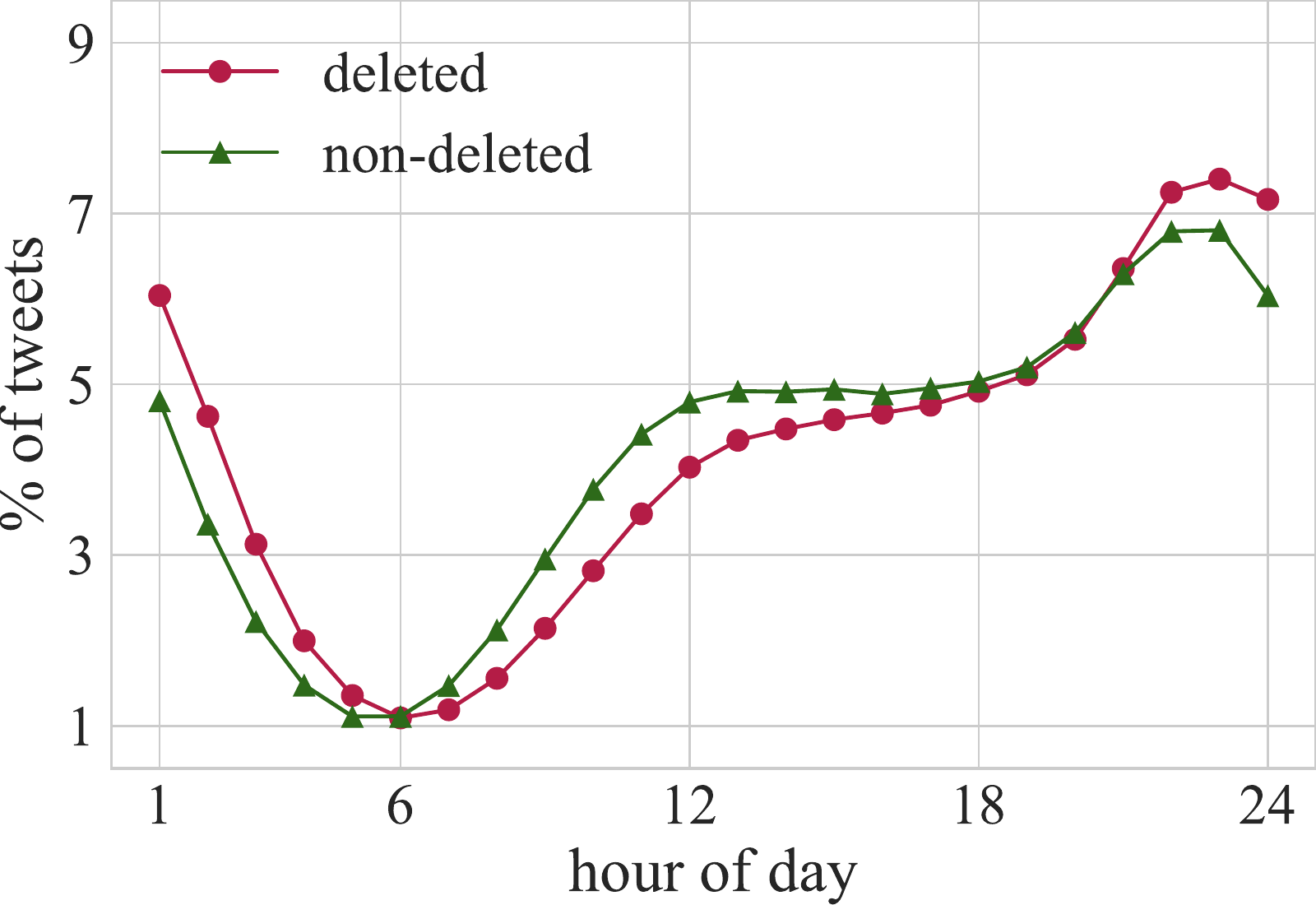}
\caption{Percentage of tweets created during every hour of the day.
Tweets that are later deleted, are more likely to be created between 8pm and 6am.}
\label{fig:good-nort-creation_hod}
\end{figure}

Earlier studies have noted that across cultures,
peoples mood show repeating patterns of variations
over daily and seasonal cycles~\cite{golder-science11}.
To understand if such patterns would also be visible
in regrettable communications on Twitter,
specifically in tweet deletion patterns,
we compared the tweet creation times of
deleted and non-deleted tweets in our dataset.
Figure~\ref{fig:good-nort-creation_hod} shows the percentage of tweets
in the deleted tweet set and the non-deleted tweet set
created during every hour of the day.
It can be observed that tweets that are later deleted
have a significantly greater probability of getting created
during the night hours in between 8pm and 6am.
Our observation is similar to that of 
Wang et al.~\shortcite{wang-cscw14}, who had observed that while
the volume of tweets with curses follows
the volume of all tweets,
cursing in Twitter reaches its peak during midnight.

\subsection{Comparing linguistic features}

Hu et al.~\shortcite{hu-icwsm13}
on comparing Twitter to other
written communication mediums
concluded that tweets are quite formal in construct.
To understand if deleted tweets
are more or less formal than non-deleted tweets,
we compared their part-of-speech distribution.
For this purpose, we used the CMU Twitter POS tagger~\cite{owoputi-naacl13}
to tag all tweets in our dataset.
For every tweet we also computed its lexical density.
Lexical density of a piece of text,
is the fraction of words in the text of types,
noun, verb, adjective, and adverb.
Lexical density is often used to measure the
terseness of a piece of text.
Further, for every tweet we computed the fraction of words
contained in it that are included in an English dictionary.
For this purpose, we used the Hunspell English dictionary%
\footnote{\url{http://hunspell.sourceforge.net/}}.

\begin{table}
\centering
\small

\begin{tabu} to \linewidth{@{}lrr@{}c@{}}
\toprule

\rowfont[c]{}
& NTD
& NUD & \\

\midrule

Proper Noun & -9.71\%  & 3.15\%   & \xitem \\
Common Noun &  2.25\%  & 11.42\%  & \\
Verb        &  3.84\%  & 136.69\% & \\
Adjective   & -0.85\%  & -27.41\% & \\
Adverb      &  3.22\%  & 54.03\%  & \\
%
\midrule

Emoticon    &  -15.00\% & -40.52\% & \\
\midrule

Lexical density & -0.70\% &	-57.57\% & \\
Dictionary words & 4.76\% &	103.74\% & \\
\bottomrule
\end{tabu}
\caption[]{
Normalized Tweet Difference (NTD) and Normalized User Difference (NUD)
with respect to different parts of speech categories,
emoticon usage, lexical density, and percentage of dictionary words used.
For proper nouns, the trend observed
from tweet (NTD) and user (NUD) perspective reverses.
The fraction of tweet vocabulary
consisting of part-of-speech tags and emoticons
(used to compute NTD)
in deleted and non-deleted tweets
have statistically significant difference
when compared using Fisher's Exact test with $p < 0.001$,
while median lexical density and median fraction of dictionary words
(used to compute NTD)
in deleted and non-deleted tweets
have statistically significant difference
when compared using Mann-Whitney U test with $p < 0.001$.}

\if{0}
  in deleted tweets and non-deleted tweets.
  Also presented is the percentage of users (Del Signif Users)
  whose deleted tweets
  have significantly more part of speech of the given kind,
  or have significantly higher lexical denisty
  or have significantly higher dictionary words,
  when compared to their non-deleted tweets.
  Similarly, the percentage of users (Non-Del Signif Users)
  for whom the reverse is true is also shown.\protect\footnotemark}
\fi
\label{tab:postag-cmp}
\end{table}

Table~\ref{tab:postag-cmp} shows the difference
in deleted and non-deleted tweets with respect to
use of different part-of-speech types,
lexical density scores, and
percentage of dictionary words in tweets.
While we find that deleted and non-deleted tweets
are similar in terms of lexical density scores,
for a larger proportion of users 
lexical density scores are significantly lower in their deleted tweets.
Also, notable is that the percentage of proper nouns
used in deleted tweets is significantly lower (9.71\% drop)
than those in non-deleted tweets.
This observation seems to also be in agreement
with the fact that deleted tweets have lower mention counts.
We postulate that this maybe because
users writing these tweets are aware of their controversial contents
and thus avoid mentioning specific people in these tweets.

We also note that there is the marked reduction of emoticons
in deleted tweets (15\% drop).
Further, while deleted tweets have a slightly higher
percentage of words from the English dictionary,
for a  large fraction of users (103.74\% more)
their deleted tweets contain significantly more dictionary words.
This taken together with the less use of emoticons,
suggest that \emph{deleted tweets are more likely to be formal}
than tweets that are not deleted.

\begin{table}
\centering

\begin{minipage}{0.4\textwidth}
\small
\centering
\begin{tabu} to \linewidth{@{}lrr@{}c@{}}
\toprule

\rowfont[c]{}
& NTD
& NUD & \\

\midrule
\multicolumn{3}{l}{\textbf{Pronouns}}       \\
1st Person (singular) & 5.82\%   & 12.22\%  & \\
1st Person (plural)   & -10.41\% & 110.74\% & \xitem \\
2nd Person            & 1.65\%   & 5.10\%   & \\
3rd Person (singular) & 10.46\%  & 279.47\% & \\
3rd Person (plural)   & 6.66\%   & 251.46\% & \\
Impersonal            & 6.50\%   & 139.73\% & \\

%

\midrule
\multicolumn{3}{l}{\textbf{Tense}} \\
Past     &  3.49\% &  62.98\% & \\
Present  & 2.41\% & -19.60\% & \xitem \\
Future   & -1.14\% & 182.35\% & \xitem \\


\midrule
\multicolumn{3}{l}{\textbf{Emotion}} \\

Positive Emotion & -13.03\% &   -69.90\%  &  \\
Negative Emotion  & 16.23\% &    269.27\% &  \\
Anxiety            & 3.70\% &    425.58\% &  \\
Anger             & 23.74\% &    566.10\% &  \\
Sadness            & 6.25\% &    218.52\% &  \\
%
%
\midrule
\multicolumn{3}{l}{\textbf{Cognitive Process}} \\

Insight     & 5.44\%&   144.68\% & \\
Causation   & 6.55\%&   241.74\% & \\
Discrepency & 5.09\%&   164.43\% & \\
Tentative   & 7.86\%&   266.43\% & \\
Certainity  & 1.49\%&    70.88\% & \\
Inhibition  & 4.34\%&   420.75\% & \\
Inclusive   & 0.40\%&    82.01\% & \\
Exclusive   & 8.48\%&   224.03\% & \\

%
\midrule
\multicolumn{3}{l}{\textbf{Swear Words}} \\
Swear Words       & 30.10\% & 691.84\% & \\
Sexual References &  7.44\% & 145.95\% & \\

\bottomrule
\end{tabu}
\end{minipage}
\hspace{0.1\textwidth}
\begin{minipage}{0.35\textwidth}
\caption[]{\small
  Normalized Tweet Difference (NTD) and Normalized User Difference (NUD)
  with respect to LIWC 2007 categroies.%

  \vspace{0.3\baselineskip}
  For 1st Person (plural) pronouns,
  present tense words, and future tense words,
  the trends are reversed when seen from the
  tweet (NTD) and user (NUD) point of view.

  \vspace{0.3\baselineskip}
  Differences in percentage
  of words belonging to different LIWC 2007 categories
  (used to compute NTD),
  in deleted and non-deleted tweets
  are significantly different
  when compared using Fisher's Exact test with $p < 0.001$.}

\if{0},
  for deleted tweets and non-deleted tweets.
  Also presented is the percentage of users (Del Signif Users)
  who use the given category of words
  signficantly more in their deleted tweets,
  when compared to their non-deleted tweets.
  Similarly, the percentage of users (Non-Del Signif Users)
  for whom the reverse is true is also shown.\protect\footnotemark}
\fi
\label{tab:liwc-cmp}
\end{minipage}
\end{table}

\Replacement{To understand the differences in language use,
we compared the deleted and non-deleted tweet corpuses
using the LIWC 2007 toolkit.
The difference in tweet vocabulary usage,
in deleted tweets and non-deleted tweets,
corresponding to the different LIWC categories,
is shown in Table~\ref{tab:liwc-cmp}.}
We find that a much higher fraction of deleted tweets
have third person pronoun usage,
while use of first person plural pronouns
decreases significantly (10.41\% drop).
Interestingly, we find
that there is a small increase
in the percentage of users who use more first person plural pronouns
in tweets they later delete.
However, the percentage of users
who use more third person singular and plural pronouns
in deleted tweets is more than 2.5 times higher.
These observations can be explained as follows.
Earlier studies on {\it ingroup outgroup bias}~\cite{brewer-jsi99}
have shown that users are more likely to talk about others,
whom they do not think of as part of their ingroup,
in a harsher language.
Note, we also observe in Table~\ref{tab:liwc-cmp}
that deleted tweets show significantly more
negative emotion than non-deleted tweets.
We postulate that the above facts may be caused
because subjects of deleted tweets
are much less likely to be viewed by the tweet's author
as part of her ingroup.

\Replacement{Additionally, we find significantly higher use
of past and present references in deleted tweets
(3.49\% increase for past tense and 2.41\% increase for present tense),
with the use of future references dropping slightly (1.14\% decrease).}
However, we  find that the normalized user difference with respect
to use of future referencing vocabulary increases to 182.35\%.
As the percentage of users who use
future tense significantly more in non-deleted tweets,
compared to their deleted tweets,
is very low
(less than 1\%) this doesn't indicate a significant trend.
Similarly, for present tense, the trend reverses when we consider user metric.
From the consistent trend exhibited in past tense and
carefully examining the tweets we find that in many of the cases deleted tweets are {\em comments about past incidents.}


\Replacement{As expected, we find that
use of words with negative emotions increase (16.23\% increase)
significantly in deleted tweets,
while a smaller percentage of deleted tweets (13.03\% drop)
have words related to positive emotions
compared to non-deleted tweets.
Additionally, we find that while there is a minor
increase in use of anxiety and sadness related words
in deleted tweets,
a significantly higher percentage of deleted tweets (23.74\% more)
contain words related to anger.}

Also notable, is that for all categories of words
related to cognitive processes,
a larger fraction of deleted tweets contain words related to them.
Similar trend is seen in the percentage of users
using significantly more cognitive process related words
in their deleted tweets compared to their non-deleted tweets.
These demonstrate that in general \emph{tweets
that are deleted in Twitter are more carefully constructed}
than general tweets in the non-deleted set.

Finally, in accordance with our observation
regarding negative emotions,
we see that a larger fraction of tweets
in the deleted tweet set contain swear words (30.10\% increase),
and words related to sexual references (7.44\% increase).
This trend is also seen in the percentage of users,
who use significantly more swear words and
sexual references in their deleted tweets.

\subsection{Summary}
Overall we find that when compared to non-deleted tweets,
deleted tweets are simultaneously less informative
and less conversational.
Deleted tweets have comparable lexical density with non-deleted tweets,
but have a significant lack of emoticon usage
and higher use of dictionary words
indicating a more formal tone.
A linguistic analysis of the two tweet sets shows that
deleted tweets are likely to be referring to an unnamed third person,
having a negative sentiment,
and have high increase in swear word usage.
Deleted tweets also show significant increase
in words related to cognitive process
indicating thoughtful construction.
In general, we find that aggregate differences in tweet features
follow the individual user differences.

\section{Comparing responses to deleted and non-deleted tweets}
\label{sec:response}

When a speaker makes a regrettable remark
in an offline conversation,
her audience's response, especially non-verbal cues,
allows her to realize it quickly and respond accordingly. 
In Twitter, replies are the primary form of feedback
that a user receives when tweeting.
However, existing studies of deleted tweets~\cite{almuhimedi-cscw13,zhou-www16}
have generally ignored this aspect of communication
when trying to understand why a tweet is being deleted.
Interestingly, Mondal et al.~\shortcite{mondal-soups16}
noted that in some cases partial content of a tweet that
has been deleted can be guessed from non-deleted tweets
that had replied or manually retweeted the original deleted tweet.
Here, we try to answer the following question:
\emph{do tweets which are deleted later,
receive significantly more negative replies,
than tweets that are not deleted?}

\begin{table}
\centering
\small

\setlength{\tabcolsep}{4pt}
\begin{tabu} to \linewidth{@{}lrr@{}}
\toprule

& Del Tweets & Non-Del Tweets\\

\midrule

Tweets w/ replies  & 15.50\% & 23.24\% \\
Tweets w/ retweets & 10.26\% & 16.14\% \\
Tweets w/ quotes   & 0.16\%  & 0.20\% \\
\bottomrule
\end{tabu}
\caption{Percentage of tweets
  that were replied to, retweeted, and quoted
  in the deleted and non-deleted tweet sets.}
\label{tab:response-stats}
\end{table}



The percentage of tweets in our dataset
receiving different forms of responses
are shown in Table~\ref{tab:response-stats}.
As we can see, deleted tweets receive significantly fewer
responses when compared to non deleted tweets.
In our dataset, 23.24\% of non-deleted tweets received replies,
while the same for deleted tweets is 15.5\%.
We also find that 16.14\% non-deleted tweets in our dataset
received one or more retweets,
while the same for deleted tweets is 10.26\%.
The trends for percentage of tweets quoted are also similar.


One may be inclined to postulate that,
the reason behind deleted tweets receiving fewer responses
is that they have a smaller lifetime
and thus lesser opportunity to receive them.
However, we find that deleted tweets
get ample opportunity to receive replies.
The median time to receive the first reply
is 128 seconds for all tweets in our dataset.
Whereas, the median time duration after which a tweet is deleted
is 1 hour and 12 minutes.
For deleted tweets which received replies in our dataset,
the median time to receive first reply was only 85 seconds.
Further, the median time duration after which such a tweet
was deleted is 5 hours and 49 minutes.

\begin{table}
\centering
\small

\begin{tabu} to \linewidth{@{}lrr@{}}
\toprule

\rowfont[c]{}
& Del Tweets & Non-Del Tweets \\

\midrule

Tweets w/ +ve reply & 63.13\% & 69.94\% \\
Tweets w/ -ve reply & 36.86\% & 30.05\% \\

\bottomrule
\end{tabu}
\caption{Percentage of tweets
  which had received replies
  with positive and negative sentiments,
  out of tweets that had received any reply,
  in the deleted and non-deleted tweet sets.}
\label{tab:response-sent-cmp}
\end{table}


Finally, to answer the question ---
whether deleted tweets receive
significantly more negative replies ---
we computed the sentiment scores for replies
to all deleted and non-deleted tweets
in our dataset.
Table~\ref{tab:response-sent-cmp}
shows percentage of deleted and non-deleted tweets
for which the first reply received
had positive and negative sentiments.
We find that a significantly lesser percentage
of deleted tweets receive positive replies
when compared to non-deleted tweets
(9.73\% drop).
While, the percentage of deleted tweets
that receive negative replies
is significantly larger
when compared to non-deleted tweets
(22.66\% increase).
Hence, it can be concluded that negative response (replies) from
the audience is correlated with deletion of tweets by the authors.

From an author's point of view, it would have been better
to have mechanisms that predict and warn against posting
of regrettable tweets, than to depend upon replies to realise the regret.
In the next section, we attempt to develop such an early
prediction system.



\section{Early detection of tweets that are deleted in future}
\label{sec:classify}

In this section, we try to exploit
the differences in user attributes and tweet features
that we observed between deleted and non-deleted tweets in the previous sections,
to build a classifier
that can predict if a tweet is likely to be deleted later.

A machine learning based system for predicting
possible tweet deletions can be utilized in two complimentary ways.
First, a system that can predict tweet deletion
from only post time features
can be useful for nudging and cautioning
its users even before they have posted a possibly regrettable tweet.
Second, after the tweet has been posted,
the system can monitor the responses that the tweet receives
and can quickly alert its author before the tweet
gets a chance to cause too much damage.
Thus, in this section we study two sets of classification tasks:
(i) predicting if a tweet is likely to be deleted
using only those features that are available at posting time,
and (ii) predicting if a tweet is likely to be deleted
from post time features as well as features obtained
from the responses received by the post.

To create a balanced dataset
for training and testing the classification process,
we randomly sampled a set of 200,000 deleted tweets
and 200,000 non-deleted tweets
\emph{from the set of tweets posted by users in the deleter set}.
We ensured that equal number of deleted and non-deleted tweets
are selected from the same user.
For the purpose of this work,
we used classifier implementations
from the scikit-learn project%
\footnote{\url{http://scikit-learn.org/}}.

\NewEnviron{MyFeatureItemize}%
{\vspace{-1.5ex}\begin{itemize}[nosep,leftmargin=1.2em]\BODY\end{itemize}}%
{}

\begin{table}
\centering
\footnotesize

\begin{tabu} to \linewidth{@{}lrX[l]@{}}
\toprule


\multirow{6}{*}{%
  \rotatebox[origin=c]{90}{%
    \colorbox{white}{%
      \parbox[c][4em][c]{34em}{
        \centering
        \colorbox{white}{\hspace{34em}}
        \colorbox{white}{\textbf{Post time features}}
        \break
        \textbf{\downbracefill}
}}}}
& \emph{Raw open text features} &
\begin{MyFeatureItemize}
\item TF-IDF score features computed using a bag of words model
\mbox{(11.7 million sparse features)}
\end{MyFeatureItemize}\\[-0.5em]

& \emph{Close text features (LIWC)} &
\begin{MyFeatureItemize}
\item LIWC scores corresponding to 64 categories (64 features)
\end{MyFeatureItemize}\\[-0.5em]

& \emph{Close text feature (sentiment)} &
\begin{MyFeatureItemize}
\item Tweet sentiment score computed using Vader toolkit
\end{MyFeatureItemize}\\[-0.5em]

& \emph{Part-of-speech features} &
\begin{MyFeatureItemize}
\item Word counts for 25 part of speech categories (25 features)
\end{MyFeatureItemize}\\[-0.5em]

& \emph{Tweet attribute features} &
\begin{MyFeatureItemize}
\item Time of the day when the tweet was posted (in UTC)
\item Day of the week when the tweet was posted (in UTC)
\item Timezone of the author posting the tweet
\item Whether the tweet is a reply to another tweet
\item Whether the tweet quotes another tweet
\item Number of urls present in the tweet
\item Number of users mentioned in the tweet
\item Number of hashtags used in the tweet
\item Whether the tweet has location information
\end{MyFeatureItemize}\\[-0.5em]

& \emph{User attribute features} &
\begin{MyFeatureItemize}
\item Creation time of the user's account
\item Whether the user has customized his profile
\item Whether the user has a custom profile image
\item Length of the bio posted by the user
\item Whether the user has enabled geo tagging of tweets
\item Whether the user has location information in his profile
\item Whether the user has a homepage url in his profile
\item Number of tweets favorited by the user
\item Number of accounts the user is following
\item Number of accounts that are following the user
\item Number of lists the user is a member of
\item Number of tweets that the user has posted
\end{MyFeatureItemize}\\[-0.5em]

\midrule

\multirow{1}{*}{%
  \rotatebox[origin=c]{90}{%
    \colorbox{white}{%
      \parbox[c][4em][c]{9em}{
        \centering
        \colorbox{white}{\textbf{Response time}}
        \colorbox{white}{\textbf{features}}
        \break
        \textbf{\downbracefill}
}}}}
& \emph{Retweet features} &
\begin{MyFeatureItemize}
\item Number of times the tweet has been retweeted
\item Number of other tweets quoting the tweet
\end{MyFeatureItemize}\\

& \emph{Reply features} &
\begin{MyFeatureItemize}
\item Number of replies the tweet has received
\item Aggregated LIWC feature vector for replies
\item Aggregated POS feature vector for replies
\item Aggregated Sentiment feature for replies
\end{MyFeatureItemize}\\

\bottomrule
\end{tabu}
\caption{Different features used for classification
of deleted and non-deleted tweets}
\label{tab:feature}
\end{table}

\subsection{Features for the classification models}

Table~\ref{tab:feature} shows the different features
used for the classification tasks.
The features are broadly divided into \emph{post time features}
and \emph{response time features}.

\emph{Raw open text features},
generated using bag of words model,
are high dimensional and sparse features.
To use them together
with other low dimensional and dense features,
we first convert them
into a single discriminative scalar feature.
This is done by training a classifier
with the raw open text features,
and using it to predict a possible deletion score for all tweets.
We call this deletion score the \emph{derived open text feature}.
Later, we use the derived open text feature
along with other dense features
to perform the final prediction.
To compute the raw open text features,
we tokenized all tweets in our dataset
using the CMU Twitter POS Tagger~\cite{owoputi-naacl13}.
After removal of mentions and urls,
this resulted in a word corpus
of 11.7 million unique terms.
Raw open text feature for a tweet
is the sparse 11.7 million dimensional tf-idf vector
corresponding to the terms of the given tweet.

The LIWC 2007 English dictionary categorizes
words into 64 different categories~\cite{liwc07}.
The \emph{Close text features (LIWC)} of a given tweet
is a 64 dimensional vector feature,
representing scores corresponding to the 64 LIWC categories.
The \emph{Close text feature (Sentiment)} of a tweet
is a scalar feature
representing the sentiment score of the tweet
computed using the Vader toolkit~\cite{hutto-icwsm14}.
Both LIWC and Vader toolkits
have a predefined fixed corpus of words (closed set),
using which they score different pieces of text
and thus produce the set of close text features.

The CMU Twitter POS Tagger~\cite{owoputi-naacl13}
distinguishes between 25 different parts of speech types.
Using it, we computed the \emph{parts-of-speech features},
representing the parts of speech distribution for every tweet.

\emph{Tweet attribute features} consist of nine features
obtained from each tweet object.
These consist of:
time of day and day of week when the tweet was posted (2 features),
time zone of the author (1 feature),
whether the tweet is a reply or a quote (2 features),
number of hashtags, urls, and user mentions in the tweet (3 features),
and whether the tweet has location information (1 feature).

\emph{User attribute features} consist of twelve features
obtained from the author information associated with the tweet.
These consist of:
age of the author's account (1 feature),
whether the user customized her profile and profile image (2 features),
length of the user's Twitter bio (1 feature),
whether the user has location and homepage url on her profile (2 features),
whether the user has enabled geotagging (1 feature),
number of followers, followees, and listed count (3 features),
and number of tweets posted and favorited by the user (2 features).

Unlike the \emph{post time features} presented above,
\emph{retweet features} and \emph{reply features}
are \emph{response time features},
and consist of attributes
which are not available at the time of posting,
and varies as the tweet accumulates more responses.
The retweet and reply features of a tweet consists of three scalar features,
corresponding to the count of other tweets,
retweeting, quoting, or replying, to the given tweet.
In addition,
we compute aggregate features
by adding up the LIWC feature vectors,
the parts-of-speech feature vectors,
and sentiment features
of the replies that the tweet has received.



\subsection{Predicting tweet deletions at posting time}

Combining millions of sparse raw open text features
and few hundreds of other dense features,
to create input for a single classifier is a difficult task.
This is because,
different types of classifiers work better
for different types of input features.
Naive Bayes and SVM with Linear kernel
are most popular classifiers in the text classification context
and can easily work with
large number of open text features~\cite{schulz-emnlp15},
while AdaBoost and SVM with RBF kernel
can capture more complex relations in data,
but are very resource expensive
when training with millions of sparse features~\cite{svm-guide}.
Thus, in this study we separately train classifiers
with only raw open text features for the purposes
of computing the \emph{derived open text feature}.
The \emph{derived open text feature}
is a scalar score produced by this first stage classifier
and is then used alongside other dense features for
the second stage classifier that utilizes all post time features.

\textbf{Predicting tweet deletions using only raw open text features:}
For predicting tweet deletions using the raw open text features,
and computing the derived open text feature in the process,
we used two classifiers well suited for the task:
(i) Multinomial Naive Bayes classifier (Naive Bayes) and
(ii) Support Vector Machine classifier with linear kernel (Linear SVM).
The hyper-parameters chosen for the two classifiers were:
$\alpha=0.1$ for the Naive Bayes classifier,
and $C=10^{-6}$ for the Linear SVM classifier with L2 regularizer.
The hyper-parameters were chosen using grid search on the parameter space.
A ten-fold cross validation was performed
for each parameter combination.
The final parameters were selected
to maximize the mean F1-Score on test sets.

\begin{table}
\centering
\small

\begin{tabu} to \linewidth{@{}lrrrr@{}}
\toprule

\rowfont[c]{}
Classifier & Precision & Recall & F1-Score \\

\midrule

Naive Bayes & 0.541 & 0.705 & 0.612 \\
Linear SVM  & 0.572 & 0.605 & 0.588 \\

\bottomrule
\end{tabu}
\caption{Classifier performance in distinguising
  deleted tweets from non-deleted tweets
  using raw open text features only.}
\label{tab:predict-open-text}
\end{table}

%
%

%
%

Table~\ref{tab:predict-open-text}
shows the performance
of the classifiers
in predicting the likelihood of a tweet being deleted.
While the Naive Bayes classifier
produces a higher F1-Score
(0.612 for Naive Bayes vs. 0.588 for Linear SVM)
due to higher recall,
its precision is significantly lower
(0.541 for Naive Bayes vs. 0.572 for Linear SVM)
when compared to the Linear SVM classifier.

Due to the higher precision and comparable recall
of the Linear SVM classifier,
we used distance of tweet features from the SVM decision boundary
as the derived open text feature.
As classifiers used in the later sections
use dense low dimensional features,
the derived open text feature serves as a dense scalar proxy
for the sparse high dimensional features.


\begin{table}
\centering
\small

\begin{tabu} to \linewidth{@{}lrrr@{}}
\toprule

\rowfont[c]{}
Classifier & Precision & Recall & F1-Score\\

\midrule

SVM      & 0.630 & 0.756 & 0.687 \\
AdaBoost & 0.796 & 0.765 & 0.780 \\


\bottomrule
\end{tabu}
\caption{Classifier performance in distinguising
  deleted tweets from non-deleted tweets
  using all post time features.}
\label{tab:predict-all}
\end{table}

%
%

%
%

\textbf{Predicting tweet deletions using all post time features:}
For predicting tweet deletions using
the derived open text feature and other dense features
obtainable at posting time,
we used two classifiers:
(i) Support Vector Machine classifier with RBF kernel (SVM) and
(ii) AdaBoost classifier with Decision Tree (AdaBoost).
The parameters chosen for the classifiers were:
$C=0.1$ with kernel coefficient $\gamma=0.001$
for the SVM classifier,
and maximum tree depth $=5$ with $100$ estimators
for the AdaBoost classifier.
As before, the hyper-parameters were chosen
using grid search on the parameter space.
A ten-fold cross validation was performed
for each parameter combination.
The final parameters were selected
to maximize the mean F1-Score on test sets.

Table~\ref{tab:predict-all} shows
that the AdaBoost classifier performs
significantly better than the SVM classifier
with a F1-Score of 0.780 compared to 0.687 for SVM.


\begin{table}
\centering
\small

\begin{tabu} to \linewidth{@{}lrrr@{}}
\toprule

\rowfont[c]{}
Dropped Feature(s) & Precision & Recall & F1-Score\\

\midrule

User features      & 81.52\% & 85.44\% & 83.46\% \\
Derived \mbox{open text feature}  & 97.29\% & 97.19\% & 97.24\% \\
Tweet features     & 99.53\% & 99.17\% & 99.34\% \\
Sentiment feature   & 99.43\% & 99.52\% & 99.47\% \\
POS features       & 99.32\% & 99.71\% & 99.51\% \\
LIWC features      & 99.67\% & 99.73\% & 99.89\% \\

\bottomrule
\end{tabu}
\caption{Relative classifier performance
  with respect to baseline AdaBoost classifier
  (Table~\ref{tab:predict-all})
  trained with all features,
  on dropping one of the sets of features
  for the task of distinguising
  deleted tweets from non-deleted tweets.}
\label{tab:feature-importance}
\end{table}







\subsection{Importance of features in predicting tweet deletions}

To understand the relative importance of
different feature groups for the classification process,
we trained AdaBoost classifiers
by dropping one group of features at a time
and comparing the resultant classifier's performance with
the baseline AdaBoost classifier (Table~\ref{tab:predict-all})
trained with all post time features.

Table~\ref{tab:feature-importance} shows that
the largest loss of performance (F1-score)
comes on removal of user features.
The second largest loss is observed
on removing the derived open text feature score.
Removal of LIWC features leads to the lowest
loss in performance.

\begin{table}
\centering
\small

\begin{tabu} to \linewidth{@{}lrrr@{}}
\toprule

\rowfont[c]{}
Classifier & Precision & Recall & F1-Score\\

\midrule

SVM & 0.709 & 0.676 & 0.692 \\
AdaBoost & 0.825 & 0.809 & 0.817 \\

\bottomrule
\end{tabu}
\caption{Classifier performance in distinguising
  deleted tweets from non-deleted tweets
  using post time as well as response features.}
\label{tab:predict-late}
\end{table}

%
%
%

%
%

\subsection{Predicting tweet deletions after posting}

To understand the utility of feedback from other users
in predicting if a tweet will be deleted or not,
we train another set of classifiers
using posting time as well as response features.
Further, we took only the subset of
deleted tweets that had received at-least one reply,
(30,152 tweets)
and an equal sized subset of non-deleted tweets
that had also received at-least one reply.
Table~\ref{tab:predict-late} shows that
using response features indeed improves classifier
performance.
The F1-Score of the AdaBoost classifier increases to 0.817,
compared to baseline (without the response features)
AdaBoost classifier with a F1-Score of 0.780.
While, the F1-Score of the SVM classifier increases to 0.692.

\begin{table}
\centering
\small

\begin{tabu} to \linewidth{@{}lrrr@{}}
\toprule

\rowfont[c]{}
Classifier & Precision & Recall & F1-Score\\

\midrule

Petrovic et al.~\shortcite{petrovic-unpub13} & 0.587 & 0.646 & 0.615 \\
Bagdouri and Oard~\shortcite{bagdouri-cikm15} & 0.604 & 0.676 & 0.638\\

\bottomrule
\end{tabu}
\caption{Performance of classifiers
  for distinguising deleted tweets from non-deleted tweets,
  presented in earlier works,
  on our dataset.}
\label{tab:predict-baseline}
\end{table}



\subsection{Differences from previous classifiers}

Preliminary works trying to classify deleted tweets
\cite{petrovic-unpub13,bagdouri-cikm15}
combined large number of sparse open text features
and tweet attribute features
into a single classifier.
This required them to use classifiers
with simple decision boundaries
that could efficiently handle such large number
of sparse open text features.
For example Petrovic et al.~\shortcite{petrovic-unpub13}
used Linear SVM,
while Bagdouri and Oard~\shortcite{bagdouri-cikm15}
utilized a Logistic Regression classifier.

To understand, how these methodologies
would perform on our dataset,
we reconstructed the classifiers proposed
in these studies using the same set
of features reported by them.
Table~\ref{tab:predict-baseline} shows
the performance of these classifiers on our dataset.
We find that our classifier,
constructed with all post time features,
performs significantly better (with F1-score 0.78)
compared to classifiers
trained using the methodologies presented by
Petrovic et al.~\shortcite{petrovic-unpub13}
and Bagdouri and Oard~\shortcite{bagdouri-cikm15}.

We posit, that the primary source
of our classifier's better performance
comes from the use of the
multi stage classification system.
Our methodology is able to incorporate classifiers
with more complex decision boundaries
(such as SVM with RBF kernel and AdaBoost classifiers)
because we fist convert the sparse and high dimensional raw open text
features into a single derived open text feature
using Linear SVM.
This allows us to utilize benefits of both sparse open text features
and powerful classifiers with complex decision boundaries.

Zhou et al.~\cite{zhou-www16} in their work
had used ten closed text features,
but limited themselves to 18\% subset of tweets,
that they defined as content-identifiable regrettable tweets.
They obtained a classification F1-score of 0.714.
However, this study was able to create a classifier
that is able to classify \emph{all} deleted and non-deleted tweets
with a F1-score of 0.78.

\subsection{Summary}

In summary, we find that
while the decision to delete a tweet
is a personal one,
classification models can be built
that can predict with significant accuracy
whether a tweet will be deleted or not.
Surprisingly, we find that user features
contribute heavily to the performance of such a model,
despite not directly containing any information about the tweet.
We hypothesize that, the reason why user features
have strong effect on classification accuracy is that
they encode significant signals about a user's personality.
We also find that responses to tweets can demonstrably improve
the prediction accuracy of such models,
indicating strong relation between
the author's choice of deleting her tweet
and the responses that tweet has received.

\section{Conclusion and discussion}
\label{sec:conclu}

The current paper
presented a large scale empirical analysis
of deleted tweets and their authors.
It utilized a series of innovations, including:
undertaking a thorough data cleaning procedure,
endeavoring to differentiate linguistic differences in
construction of deleted and non-deleted tweets,
and attempting to infer personality based correlations
pertaining to post deletion behavior of Twitter users.
This study also presented
the design of a two-stage classifier
for predicting if a tweet will be deleted or not.
The classifier utilized open-text features, closed text features,
user features, tweet features, as well as contextual features
to perform the above prediction task.

We developed several insights during the course of our study.
In particular, we found that there exists
significant differences in personality
between users who delete their tweets (even low numbers)
and those who do not.
Users who delete their tweets
are more likely to be
extroverted and neurotic
while also being less conscientious.
We also found that vocabulary of tweets,
that are later deleted,
contain significantly higher fraction of swear words
and markers indicating anger, anxiety, and sadness.
Interestingly, a significant proportion of them were written
in a more formal tone.
Surprisingly, when manually annotating tweets
to understand their causes of deletion,
we found that causes of deletion of regrettable tweets
could be only guessed for 16\% of the deleted tweets.
In Twitter, replies are the primary method
for users to respond to a tweet's author.
We found that replies to tweets that are deleted later,
have higher negative sentiments,
which prompts us to postulate that
authors of the original tweets
take these responses into account
when choosing to delete their tweets.
Finally, using the insights obtained from the analysis
of deleted tweets and their users,
we built a classification model
which can predict, at the time of posting,
if a tweet will be deleted or not,
with a F1-score of 0.78.
Further, F1-score of the classification model increases to 0.81,
when features from responses to the original tweets
are included.

One major source of tweet deletions (14\%)
in our dataset turned out to be superficial deletions,
that is the tweets were deleted by their authors
to post a fixed or slightly modified version of the original tweet.
Twitter as of date doesn't provide a method to edit tweets.
The result of which is that
any user trying to edit any of her tweets
needs to first delete the offending tweet
and compose a new one.
While there has been considerable demand for editing
facility from celebrities and popular media,%
\footnote{\url{http://www.wired.com/2013/04/what-twitter-needs/}}
allowing users to edit tweets definitely has its pros and cons.
On one hand,
deleting and re-composing the tweet is often
a problem for users
as the process severs the links
between the tweet and its responses.
On the other hand,
allowing users to edit tweets
can produce numerous forms of gaming systems
where a user of a highly retweeted tweet can
change the content of the original tweet
which may not be in agreement with the users who
did the retweeting.

An obvious next future work would be
to develop an online system
that can nudge users into making better judgments,
of whether to post a tweet or not.
A practical challenge in doing that, however,
would be in collection of enough personal deleted data
to learn the class of tweets that
the user is likely to regret and delete later.
Earlier works~\cite{petrovic-unpub00,bagdouri-cikm15} trying to predict the
possibility of a tweet getting deleted
focused mostly on the content of the tweet itself.
However, as we note in the current study,
whether the author of a tweet finds a tweet to be regrettable or not,
depends deeply on the author herself.
Thus, it is imperative for developers,
trying to create systems that nudge users towards better choices,
to ensure that the classifiers they develop are personalized
and take into account personality and behavior of the individual user
when making such predictions.

An even bigger challenge, however,
for developers of nudging systems such as the one described above,
would be to find a balance,
such that while being useful and delivering appropriate nudges,
the system refrains from actively nagging and antagonizing
its users.
Wang et al.~\cite{wang-chi14} in their study
to nudge users towards better usage of privacy features
noted that
about half the users in their survey
held negative or indifferent attitudes
towards the widgets presented to help them.
A similar response could be seen
in users of Google Mail Goggles,%
\footnote{%
Mail Goggles~\cite{mail-goggles} was an optional widget
available with Google Mail during October 2008 to May 2012.
When enabled, Mail Goggle users
were required to answer simple math questions
to demonstrate cognitive abilities
before it allowed them to send emails.
The service was intended to stop drunk email late at night.}
many of whom found it annoying rather than helpful.
We hypothesize that whether a user finds it useful
to trade off ease of use for increased privacy,
depends heavily on the personality
of the user as well as the importance the user places
on the requirement of privacy.
Thus, it is important for system designers
to not only understand the personality of the users
being exposed to such systems
but also to educate and convince them
of the problems their system is designed to solve.
Otherwise, systems of these categories
run the risk of being mostly ignored.

Another interesting and unanswered question that remains,
is that if a tweet has received significant negative response,
what should be done with?
Mondal et al.~\cite{mondal-soups16} noted
that even when deleted,
much of the content of tweets can be guessed
from the non-deleted responses that remain,
which leads potentially significant privacy leaks.

In our future studies, we intend to investigate the above issues.

\bibliographystyle{abbrv}
\bibliography{pbrefs,pbwebrefs}

\end{document}